\documentclass[twoside,twocolumn,9pt]{article}
\usepackage{extsizes}
\usepackage[super,sort&compress,comma]{natbib} 
\usepackage[version=3]{mhchem}
\usepackage[left=1.5cm, right=1.5cm, top=1.785cm, bottom=2.0cm]{geometry}
\usepackage{balance}
\usepackage{widetext}
\usepackage{times,mathptmx}
\usepackage{sectsty}
\usepackage{graphicx} 
\usepackage{lastpage}
\usepackage[format=plain,justification=raggedright,singlelinecheck=false,font={stretch=1.125,small,sf},labelfont=bf,labelsep=space]{caption}
\usepackage{float}
\usepackage{fancyhdr}
\usepackage{fnpos}
\usepackage[english]{babel}
\usepackage{array}
\usepackage{droidsans}
\usepackage{charter}
\usepackage[T1]{fontenc}
\usepackage[usenames,dvipsnames]{xcolor}
\usepackage{setspace}
\usepackage[compact]{titlesec}

\definecolor{cream}{RGB}{222,217,201}

\begin{document}

\pagestyle{fancy}
\thispagestyle{plain}
\fancypagestyle{plain}{

\fancyhead[C]{\includegraphics[width=18.5cm]{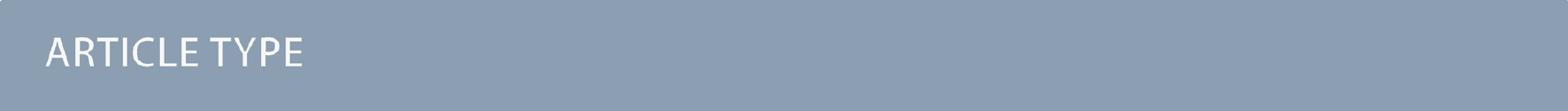}}
\fancyhead[L]{\hspace{0cm}\vspace{1.5cm}\includegraphics[height=30pt]{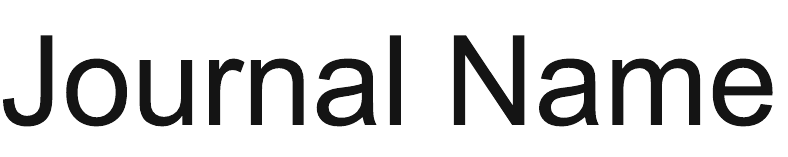}}
\fancyhead[R]{\hspace{0cm}\vspace{1.7cm}\includegraphics[height=55pt]{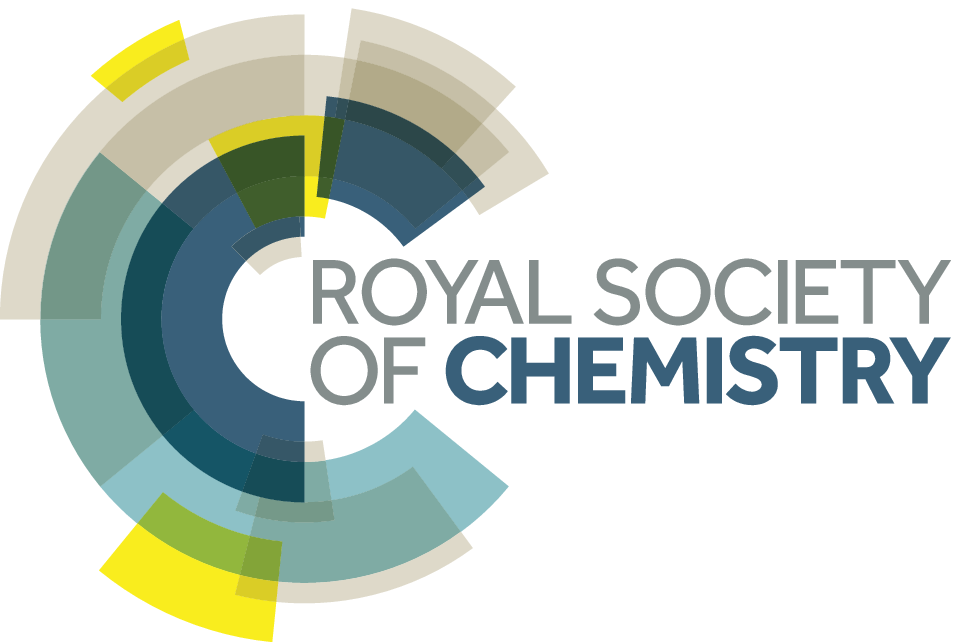}}
\renewcommand{\headrulewidth}{0pt}
}

\makeFNbottom
\makeatletter
\renewcommand\LARGE{\@setfontsize\LARGE{15pt}{17}}
\renewcommand\Large{\@setfontsize\Large{12pt}{14}}
\renewcommand\large{\@setfontsize\large{10pt}{12}}
\renewcommand\footnotesize{\@setfontsize\footnotesize{7pt}{10}}
\makeatother

\renewcommand{\thefootnote}{\fnsymbol{footnote}}
\renewcommand\footnoterule{\vspace*{1pt}%
\color{cream}\hrule width 3.5in height 0.4pt \color{black}\vspace*{5pt}} 
\setcounter{secnumdepth}{5}

\makeatletter 
\renewcommand\@biblabel[1]{#1}            
\renewcommand\@makefntext[1]%
{\noindent\makebox[0pt][r]{\@thefnmark\,}#1}
\makeatother 
\renewcommand{\figurename}{\small{Fig.}~}
\sectionfont{\sffamily\Large}
\subsectionfont{\normalsize}
\subsubsectionfont{\bf}
\setstretch{1.125} 
\setlength{\skip\footins}{0.8cm}
\setlength{\footnotesep}{0.25cm}
\setlength{\jot}{10pt}
\titlespacing*{\section}{0pt}{4pt}{4pt}
\titlespacing*{\subsection}{0pt}{15pt}{1pt}

\fancyfoot{}
\fancyfoot[LO,RE]{\vspace{-7.1pt}\includegraphics[height=9pt]{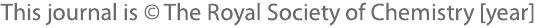}}
\fancyfoot[CO]{\vspace{-7.1pt}\hspace{13.2cm}\includegraphics{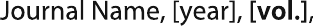}}
\fancyfoot[CE]{\vspace{-7.2pt}\hspace{-14.2cm}\includegraphics{head_foot/RF}}
\fancyfoot[RO]{\footnotesize{\sffamily{1--\pageref{LastPage} ~\textbar  \hspace{2pt}\thepage}}}
\fancyfoot[LE]{\footnotesize{\sffamily{\thepage~\textbar\hspace{3.45cm} 1--\pageref{LastPage}}}}
\fancyhead{}
\renewcommand{\headrulewidth}{0pt} 
\renewcommand{\footrulewidth}{0pt}
\setlength{\arrayrulewidth}{1pt}
\setlength{\columnsep}{6.5mm}
\setlength\bibsep{1pt}

\makeatletter 
\newlength{\figrulesep} 
\setlength{\figrulesep}{0.5\textfloatsep} 

\newcommand{\topfigrule}{\vspace*{-1pt}%
\noindent{\color{cream}\rule[-\figrulesep]{\columnwidth}{1.5pt}} }

\newcommand{\botfigrule}{\vspace*{-2pt}%
\noindent{\color{cream}\rule[\figrulesep]{\columnwidth}{1.5pt}} }

\newcommand{\dblfigrule}{\vspace*{-1pt}%
\noindent{\color{cream}\rule[-\figrulesep]{\textwidth}{1.5pt}} }

\makeatother

\twocolumn[
  \begin{@twocolumnfalse}
\vspace{3cm}
\sffamily
\begin{tabular}{m{4.5cm} p{13.5cm} }

\includegraphics{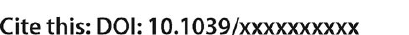} & \noindent\LARGE{\textbf{Orientation of topological defects in 2D nematic liquid crystals}} \\
\vspace{0.3cm} & \vspace{0.3cm} \\

 & \noindent\large{Xingzhou Tang\textit{$^{a}$} and Jonathan V. Selinger$^{\ast}$\textit{$^{a}$}} \\

\includegraphics{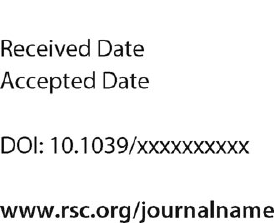} & \noindent\normalsize{Topological defects are an essential part of the structure and dynamics of all liquid crystals, and they are particularly important in experiments and simulations on active liquid crystals.  In a recent paper, Vromans and Giomi [\textit{Soft Matter}, 2016, \textbf{12}, 6490] pointed out that topological defects are not point-like objects but actually have orientational properties, which strongly affect the energetics and motion of the defects.  That paper developed a mathematical formalism which describes the orientational properties as vectors.  Here, we agree with the basic concept of defect orientation, but we suggest an alternative mathematical formalism.  We represent the defect orientation by a tensor, with a rank that depends on the topological charge:  rank 1 for a charge of $+1/2$, rank 3 for a charge of $-1/2$.  Using this tensor formalism, we calculate the orientation-dependent interaction between defects, and we present numerical simulations of defect motion.} \\

\end{tabular}

 \end{@twocolumnfalse} \vspace{0.6cm}

  ]

\renewcommand*\rmdefault{bch}\normalfont\upshape
\rmfamily
\section*{}
\vspace{-1cm}


\footnotetext{\textit{$^{a}$~Liquid Crystal Institute, Kent State University, Kent, OH 44242, USA; E-mail: jselinge@kent.edu}}





\section{Introduction}

Topological defects are common in many areas of physics, from high-energy physics and cosmology to crystal structure and superconductivity.\cite{Chaikin1995,Kleman2003}  Indeed, the importance of topological defects was recognized by the 2016 Nobel Prize in Physics.  In the science of liquid crystals, topological defects are often used to recognize phases, and they determine the structure and dynamics of many liquid-crystal phases with technological applications.\cite{deGennes1993}

In recent years, topological defects have particularly been studied in the context of two-dimensional (2D) \textit{active} nematic liquid crystals,\cite{Marchetti2013} which typically occur in systems of rod-like macromolecules on a surface.  This research has identified two new features of topological defects.  First, in active systems, defects of topological charge $+1/2$ are constantly moving, with motion driven by the activity of the material.  By contrast, defects of topological charge $-1/2$ are almost at rest, with only slow diffusive motion.  Second, both $+1/2$ and $-1/2$ defects have characteristic orientations.  For $+1/2$ defects, the orientation can been seen in the comet-like texture of the director field, and it is also the direction of the driven motion.  For $-1/2$ defects, the orientation can be seen in the three-fold symmetric texture of the director field around the defect.  Surprisingly, experiments and simulations have both shown that active systems can have long-range order in the orientation of the defects.\cite{DeCamp2015,Bartolo2015,Oza2016}  The defects tend to maintain alignment along some spontaneously chosen axis, even as they are constantly created and annihilated.

In order to understand this long-range order, one must consider the general concept of defect orientation.  A geometric question is:  How can the orientation of a defect be described mathematically?  Further physical questions are:  How does the orientation affect the interaction between two defects, or the motion of a defect?  Although these questions are motivated by studies of active nematic systems, they are relevant to all nematic liquid crystals, even equilibrium phases.

In the context of 3D nematic liquid crystals, \v{C}opar et al.\cite{Copar2011,Copar2014}\ investigate defect orientation by defining the splay-bend parameter.  This parameter provides an excellent method to visualize the orientation of a disclination line, but it does not give a mathematical expression for the orientation.  More recently, in the context of 2D nematic liquid crystals, Vromans and Giomi\cite{Vromans2016} argue that the orientational properties of nematic defects can be described by vectors, and give an explicit expression for the vectors.  Based on the vector construction, they calculate the orientation-dependent interaction between defects.  Furthermore, they model the relaxational dynamics of interacting defects, and find that the trajectory depends strongly on the defect orientation.

The purpose of our current work is to examine the concept of defect orientation in 2D nematic liquid crystals in more detail.  Through this study, we partially agree and partially disagree with the work of Vromans and Giomi.  First, in Sec.~2, we show that their vector formalism is quite reasonable for $+1/2$ defects.  However, for other defect charges, we represent the defect orientation by a tensor, with a rank that depends on the topological charge of the defect and on the symmetry of the underlying phase.  In a 2D nematic phase, a $+1/2$ defect is represented by a tensor of rank 1 (i.e. a vector), while a $-1/2$ defect is represented by a tensor of rank 3.  For a general $n$-atic phase (with an orientational order parameter of $n$-fold symmetry), a topological defect of charge $k$ is represented by a tensor of rank $n|1-k|$.

Next, in Sec.~3, we investigate the interaction between neighboring defects.  Using conformal mapping, we determine the director field around two defects with arbitrary orientations, and calculate the elastic energy associated with that director field.  The result is different from the interaction reported by Vromans and Giomi.  In Sec.~4, we construct partial differential equations to model the relaxational dynamics of a 2D nematic phase, and use these equations to simulate the annihilation of $\pm1/2$ defects with arbitrary initial orientations.  For the dynamics, we agree with the results of Vromans and Giomi:  The defect trajectories are quite similar to those reported in their paper, and these trajectories depend sensitively on defect orientation.  Finally, in Sec.~5, we discuss defect orientation as a concept for understanding the physics of 2D nematic liquid crystals.

\section{Orientation and tensor structure of defects}

Like Vromans and Giomi, we consider a 2D nematic liquid crystal.  This phase has orientational order of the molecules along the local director $\mathbf{n}(\mathbf{r})=(\cos\theta(\mathbf{r}),\sin\theta(\mathbf{r}))$.  Because the orientational order is two-fold symmetric, the director $\mathbf{n}$ is equivalent to $-\mathbf{n}$, and hence $\theta$ is only defined modulo $\pi$.  For that reason, the topological defects in this phase are disclination points, about which the director rotates through a multiple of $\pi$, so that $\oint d\theta=2\pi k$.  Here, $k$ is the topological charge of the defect, which must be an integer or half-integer, positive or negative.

\begin{figure}
\centering
  \begin{tabular*}{0.5\textwidth}{@{\extracolsep{\fill}}cc}
    \includegraphics[width=4.2cm]{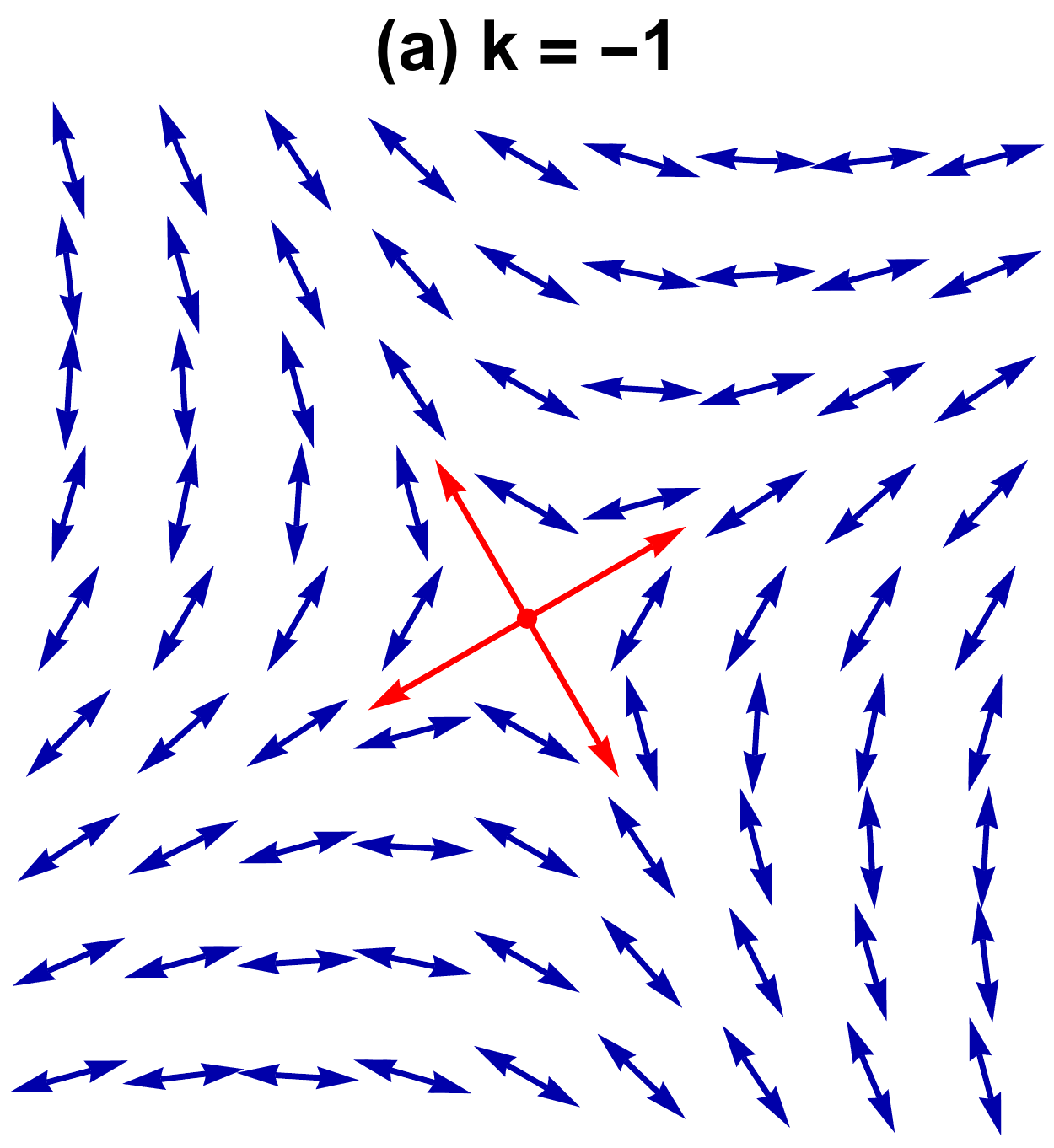} & \includegraphics[width=4.2cm]{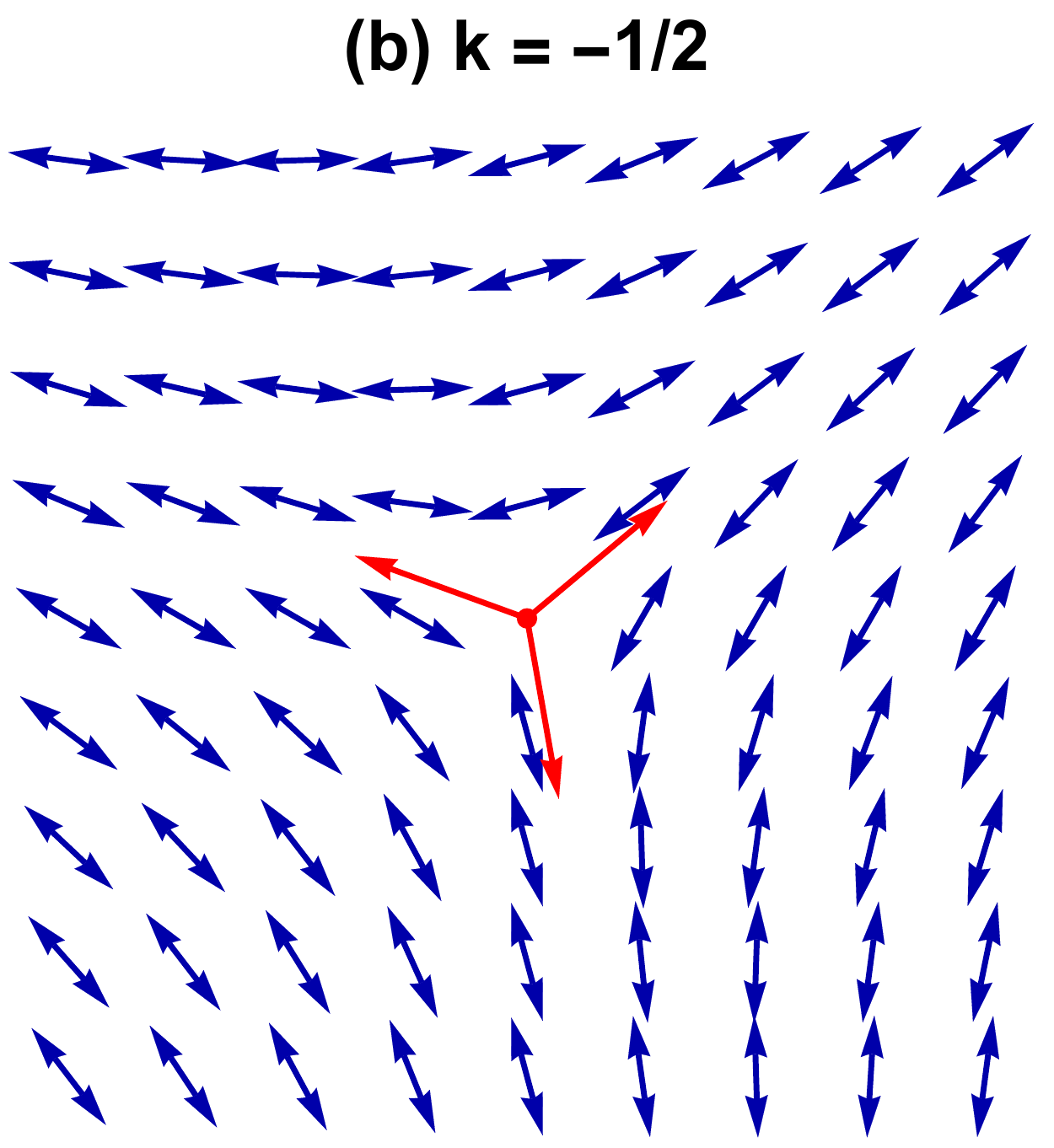} \\
    \includegraphics[width=4.2cm]{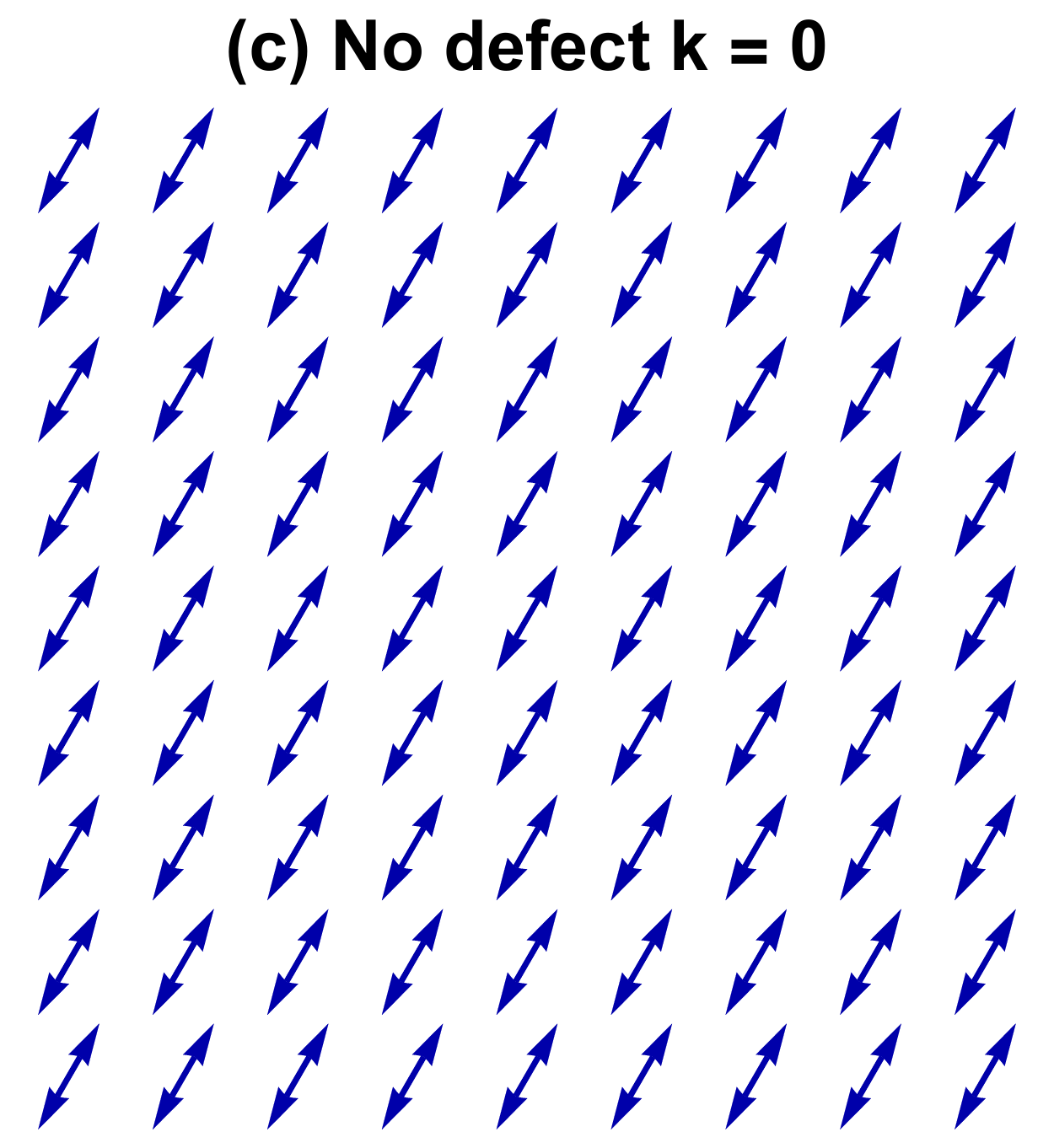} & \includegraphics[width=4.2cm]{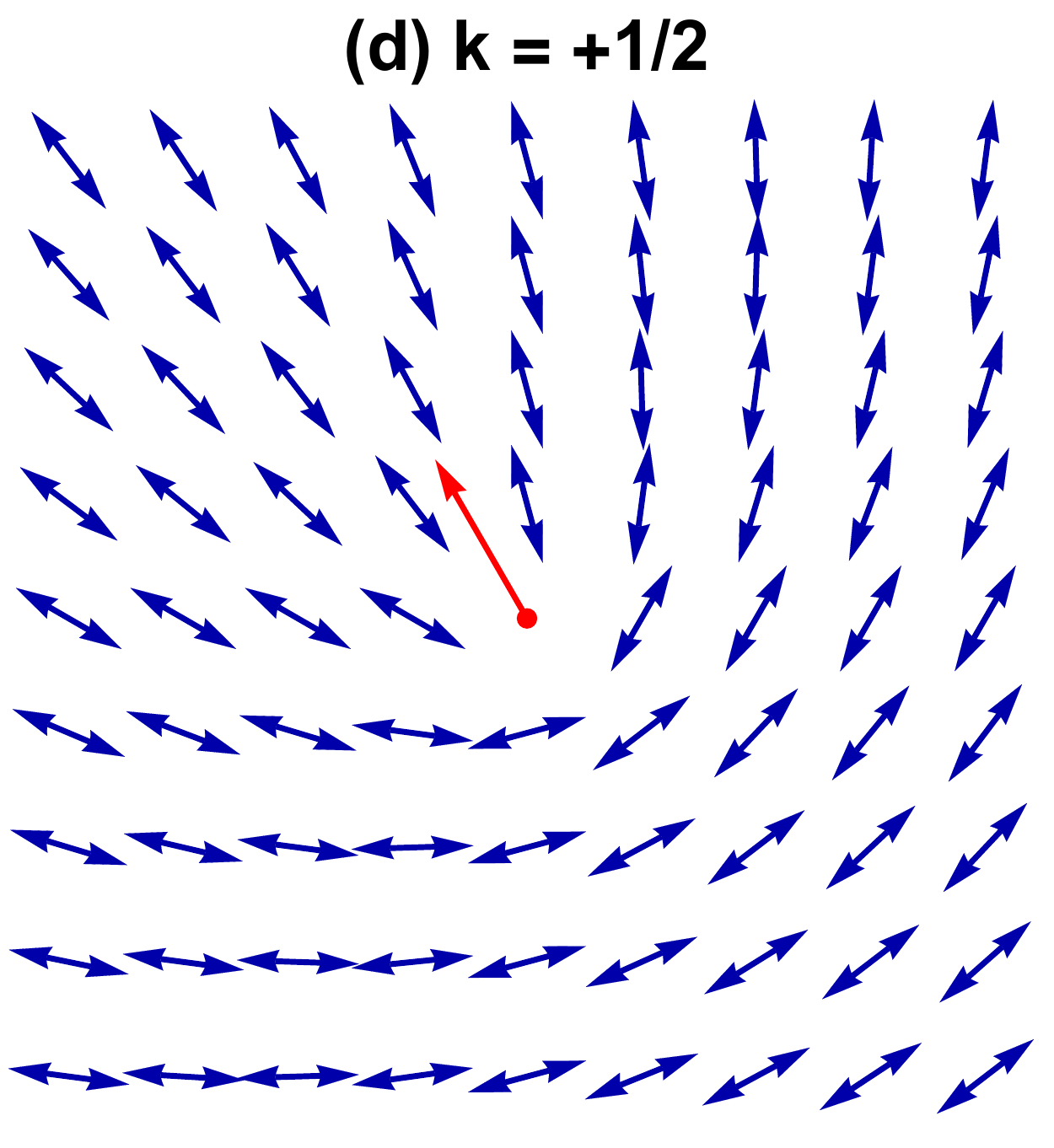} \\
    \includegraphics[width=4.2cm]{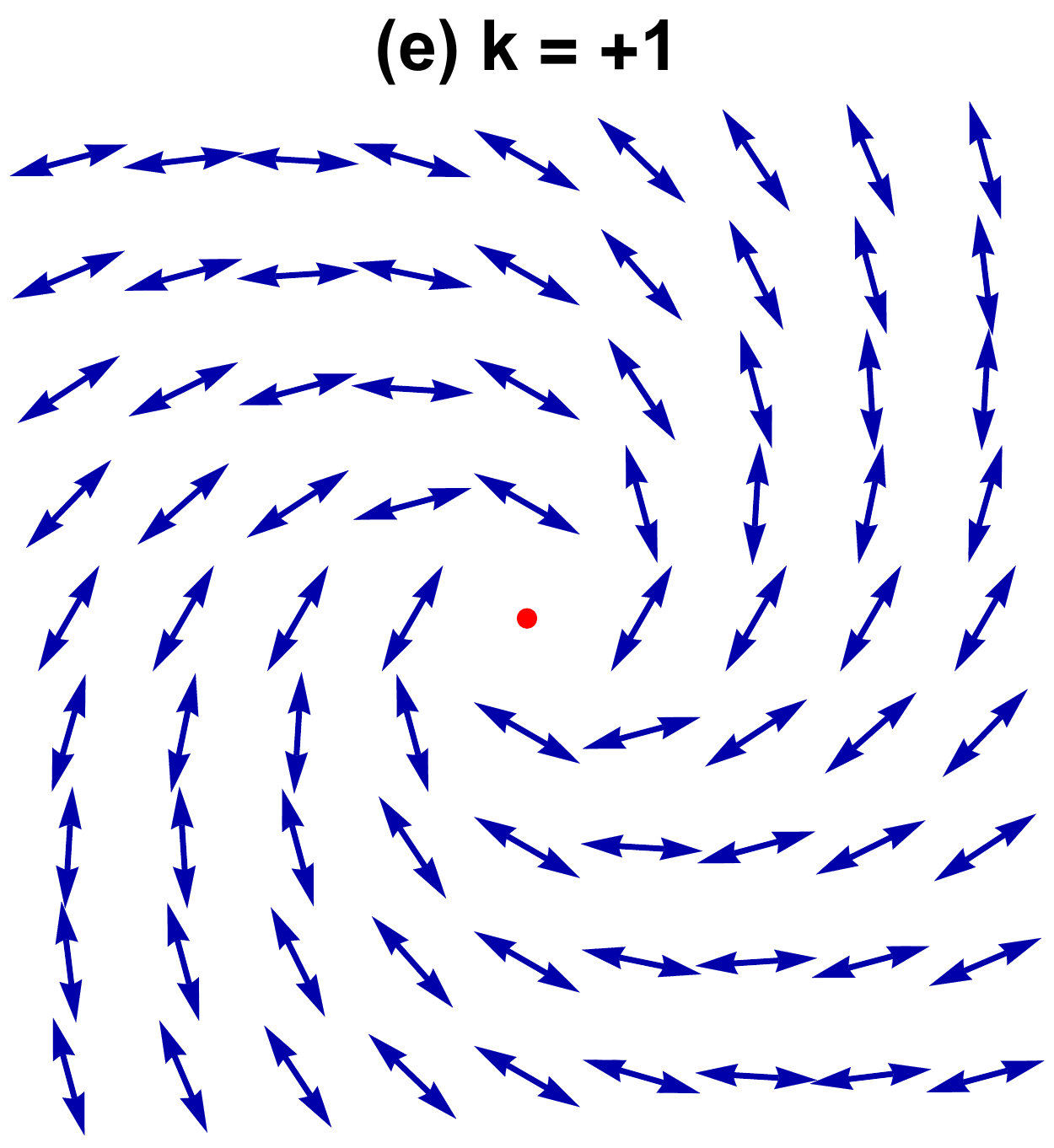} & \includegraphics[width=4.2cm]{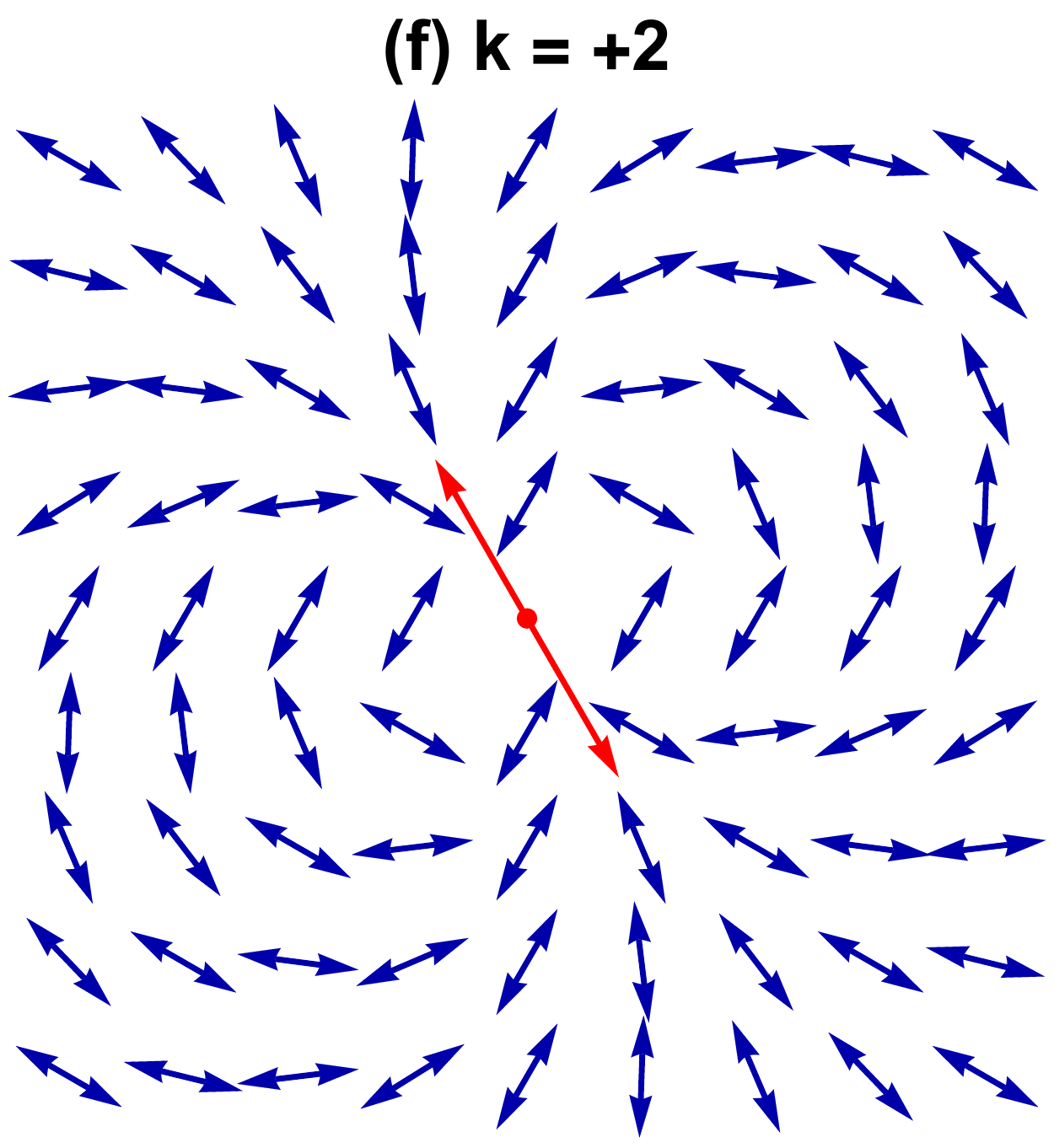} \\
  \end{tabular*}
  \caption{Examples of defects in a 2D nematic liquid crystal, with red arrows indicating the defect orientation.}
  \label{nematicdefects}
\end{figure}

The elastic free energy of the 2D nematic phase is the Frank free energy.  In the approximation of equal Frank elastic constants, this free energy can be written as
\begin{equation}
F=\frac{K}{2} \int d^2 r |\nabla\theta|^2,
\label{frankfreeenergy}
\end{equation}
where $K$ is the single Frank constant.  The local minimum of $F$ corresponding to a defect of topological charge $k$ at the origin is given by
\begin{equation}
\theta=k \phi + \theta_0,
\label{thetaarounddefect}
\end{equation}
where $\phi=\tan^{-1} (y/x)$ is the angle in polar coordinates.  The angle $\theta_0$ represents an arbitrary overall rotation of the director about the $z$-axis.  Figure~\ref{nematicdefects} shows several examples of these defects, with $\theta_0 = \pi/3$ in all cases.

To characterize the orientation of a defect, we can ask:  Where does the director $\mathbf{n}$ point radially outward from (or inward toward) the defect?  This occurs when the angle $\theta$ satisfies
\begin{equation}
\theta=\phi\quad(\text{mod }\pi).
\label{thetaequalsphi}
\end{equation}
Solving Eqs.~(\ref{thetaarounddefect}) and~(\ref{thetaequalsphi}) simultaneously, we find that these special radial directions are given by $\theta=\phi=\psi$, where
\begin{equation}
\psi=\frac{\theta_0}{1-k} \quad\left( \text{mod } \frac{\pi}{|1-k|} \right).
\label{psidefinition}
\end{equation}
Hence, the radial directions can be described by the vector $\mathbf{p}=(\cos\psi,\sin\psi)$, which is defined up to rotations through $\pi/|1-k|$.  This is precisely the defect orientation vector defined by Vromans and Giomi.  In Fig.~\ref{nematicdefects}, the red arrows show the rotationally equivalent $\mathbf{p}$ vectors for each topological charge $k$.

We now must consider how the $\mathbf{p}$ vector is related to the director field around the defect, for different values of $k$.

\subsection{Topological charge $k=+1/2$}

For topological charge $k=+1/2$, Eq.~(\ref{psidefinition}) shows that the angle $\psi$ is defined modulo $2\pi$.  As a result, the defect orientation vector $\mathbf{p}=(\cos\psi,\sin\psi)$ is a single-valued vector, as shown in Fig.~\ref{nematicdefects}(d).  It should be possible to determine $\mathbf{p}$ from the director field around the defect, and conversely, to determine the director field around the defect from $\mathbf{p}$.

To determine $\mathbf{p}$ from the director field, Vromans and Giomi show that
\begin{equation}
\mathbf{p}=\frac{\nabla\cdot(\mathbf{n n})}{|\nabla\cdot(\mathbf{n n})|}.
\label{pforplushalf}
\end{equation}
To determine the director field around the defect, we construct a covariant expression for the tensor $\mathbf{n n}$ in terms of the defect orientation vector $\mathbf{p}$ and the position $\mathbf{r}$,
\begin{equation}
n_i n_j = \frac{\delta_{ij}}{2} 
+ \frac{r_i p_j + r_j p_i - (\mathbf{r}\cdot\mathbf{p})\delta_{ij}}{2|\mathbf{r}|} .
\label{plushalfcovariant}
\end{equation}
An explicit calculation shows that this expression gives the same director field as Eq.~(\ref{thetaarounddefect}) with $k=1/2$.  As a specific check, we can see the behavior in the $\mathbf{p}$ direction out from the defect:  If $\mathbf{r}/|\mathbf{r}|=\mathbf{p}$, then $n_i n_j = p_i p_j$, so that the director field points radially outward (or inward).  As another check, the divergence of Eq.~(\ref{plushalfcovariant}) is $\partial_i (n_i n_j) = p_j /(2|\mathbf{r}|)$, and hence the vector $\mathbf{p}$ is the normalized version of $\nabla\cdot(\mathbf{n n})$, consistent with Eq.~(\ref{pforplushalf}).   Hence, this covariant expression provides a way to work with the director field in terms of $\mathbf{p}$ and $\mathbf{r}$.

Because the covariant expression of Eq.~(\ref{plushalfcovariant}) is equivalent to Eq.~(\ref{thetaarounddefect}) with $k=1/2$, it is valid in the same regime:  outside a core radius $r_\text{core}$, up to the distance where the director field is affected by different mechanisms beyond the central defect, such as other defects or boundaries.  These equations are derived with the approximation of equal Frank elastic constants.  If the Frank constants are unequal, then the director field would have corrections with higher powers of $\mathbf{r}/|\mathbf{r}|$.

As Vromans and Giomi point out, one physical interpretation of Eq.~(\ref{pforplushalf}) is in the context of active nematic liquid crystals:  The stress tensor includes an active term proportional to $\mathbf{n n}$, with positive or negative sign depending on whether the material is contractile or extensile.  The active force on the defect is then in the direction $\nabla\cdot(\mathbf{n n})$.  Hence, $\mathbf{p}$ gives the direction of defect motion in an active nematic.

We can suggest another physical interpretation in the context of flexoelectricity.\cite{Meyer1969,Buka2013}  The divergence term can be expanded as
\begin{equation}
\nabla\cdot(\mathbf{n n})=\mathbf{n}(\nabla\cdot\mathbf{n})+(\mathbf{n}\cdot\nabla)\mathbf{n}
=\mathbf{n}(\nabla\cdot\mathbf{n})-\mathbf{n}\times(\nabla\times\mathbf{n}).
\end{equation}
The right side of this equation is the standard expression for the flexoelectric polarization of a liquid crystal, in the case where the splay and bend flexoelectric coefficients $e_1$ and $e_3$ are equal (which is an approximation similar to the approximation of equal Frank constants).  Hence, a $+1/2$ defect is a point of highly concentrated flexoelectric polarization, in the direction given by $\mathbf{p}$.  As discussed by \v{C}opar et al.,\cite{Copar2011,Copar2014} the splay-bend parameter is the divergence of this flexoelectric polarization.

\subsection{Topological charge $k=-1/2$}

For topological charge $k=-1/2$, Eq.~(\ref{psidefinition}) shows that the angle $\psi$ is defined modulo $2\pi/3$.  As a result, the defect orientation vector $\mathbf{p}=(\cos\psi,\sin\psi)$ is really a triple-valued vector, as shown in Fig.~\ref{nematicdefects}(b).  In other words, three vectors are equivalent representations of the same defect orientation:  $\mathbf{p}^{(1)}$, $\mathbf{p}^{(2)}$, and $\mathbf{p}^{(3)}$, which are related to each other by rotations through $2\pi/3$.  This triple-valued vector is not necessarily a problem; physicists often need to work with mathematical objects that are multiple-valued.  (Indeed, the director field $\mathbf{n}$ is an example of a double-valued vector field, because $\mathbf{n}$ and $-\mathbf{n}$ are equivalent representations of the same orientational distribution.)

Vromans and Giomi extract $\mathbf{p}$ from the director field around a $-1/2$ defect through a two-step derivation:  They first define an intermediate vector $\mathbf{p}'$, which depends on the arbitrary choice of coordinate system, and then use $\mathbf{p}'$ to calculate $\mathbf{p}$, which does not depend on the coordinate system.  Here, we propose an alternative approach, which is consistent with their formalism but may be clearer because it explicitly uses the symmetry of the system.

Our alternative approach is based on a higher-rank tensor.  In general, an object with $r$-fold symmetry can be represented by a single-valued, completely symmetric tensor of rank $r$.  Hence, for a $-1/2$ defect, we consider the tensor $\mathbf{T}$ with components
\begin{equation}
T_{ijk}=\frac{2}{3}\left(p^{(1)}_i p^{(1)}_j p^{(1)}_k + p^{(2)}_i p^{(2)}_j p^{(2)}_k + p^{(3)}_i p^{(3)}_j p^{(3)}_k \right).
\label{tfromp}
\end{equation}
This tensor is invariant under rotations of $2\pi/3$, and it is normalized so that
\begin{equation}
|\mathbf{T}|^2 \equiv T_{ijk} T_{ijk} = 1 .
\end{equation}
To determine the director field around the defect, we construct a covariant expression for the tensor $\mathbf{nn}$ in terms of the defect orientation tensor $\mathbf{T}$ and the position $\mathbf{r}$,
\begin{equation}
n_i n_j = \frac{\delta_{ij}}{2} + \frac{T_{ijk} r_k}{|\mathbf{r}|}.
\label{minushalfcovariant}
\end{equation}
An explicit calculation shows that this expression gives the same director field as Eq.~(\ref{thetaarounddefect}) with $k=-1/2$.  As a specific check, we can see the behavior in the $\mathbf{p}^{(1)}$ direction out from the defect:  If $\mathbf{r}/|\mathbf{r}|=\mathbf{p}^{(1)}$, then $n_i n_j = p^{(1)}_i p^{(1)}_j$, so that the director field points radially outward (or inward).  The same is true for $\mathbf{p}^{(2)}$ and $\mathbf{p}^{(3)}$.

To determine the defect orientation tensor from the director field, we take the gradient of Eq.~(\ref{minushalfcovariant}),
\begin{equation}
\partial_k (n_i n_j) = \frac{1}{|\mathbf{r}|}\left[T_{ijk}-\frac{T_{ijl} r_l r_k}{|\mathbf{r}|^2}\right].
\end{equation}
By averaging this expression in a region around the defect, so that $\langle r_l r_k /|\mathbf{r}|^2\rangle = \delta_{lk}/2$, we obtain
\begin{equation}
\langle\partial_k (n_i n_j)\rangle = \frac{1}{2|\mathbf{r}|}T_{ijk}.
\end{equation}
Hence, the defect orientation tensor can be determined from the director field as the normalized average
\begin{equation}
\mathbf{T}=\frac{\langle\nabla(\mathbf{n n})\rangle}{|\langle\nabla(\mathbf{n n})\rangle|}.
\label{tensorfromdirectora}
\end{equation}
A more symmetric version of this expression is
\begin{align}
\label{tensorfromdirectorb}
&T_{ijk}=\\
&\frac{\langle\partial_i(n_j n_k)+\partial_j(n_k n_i)+\partial_k(n_i n_j)\rangle}{\sqrt{\langle\partial_a(n_b n_c)+\partial_b(n_c n_a)+\partial_c(n_a n_b)\rangle\langle\partial_a(n_b n_c)+\partial_b(n_c n_a)+\partial_c(n_a n_b)\rangle}},\nonumber
\end{align}
Equations~(\ref{tensorfromdirectora}) and~(\ref{tensorfromdirectorb}) are equivalent for the exact director field of Eq.~(\ref{minushalfcovariant}), but Eq.~(\ref{tensorfromdirectorb}) might be more suitable for numerical calculations in general.  Once we have $T_{ijk}$, we can determine the three directions $\mathbf{p}^{(1)}$, $\mathbf{p}^{(2)}$, and $\mathbf{p}^{(3)}$ through the following construction:  Define a test unit vector $\mathbf{b}=(\cos\beta,\sin\beta)$, calculate the scalar $f(\beta)=T_{ijk} b_i b_j b_k =\frac{1}{2}\cos3(\beta-\psi)$, and find the maxima of $f(\beta)$.  Those maxima occur at the angles $\beta=\psi$ (mod $2\pi/3$).  This construction identifies the principal orientation of the third-rank tensor $T_{ijk}$, just as an eigenvector calculation identifies the principal axes of a second-rank tensor.  In that sense, it can be considered as a generalized eigenvector calculation.

We have tested this construction on sample textures containing $-1/2$ defects, and confirm that it identifies the three directions where the director field points outward (or inward).  Hence, it identifies the orientation of the defects, in a way that does not depend on any choice of coordinate system.

\subsection{Other topological charges in a 2D nematic phase}

For a defect with an arbitrary topological charge $k$, Eq.~(\ref{psidefinition}) shows that the angle $\psi$ is defined up to rotations through $\pi/|1-k|$.  Hence, the defect has $2|1-k|$-fold rotational symmetry, with an orientation represented equivalently by $2|1-k|$ distinct $\mathbf{p}$ vectors.  Instead of this multiple-valued vector, we can also describe it by a single-valued, completely symmetric tensor of rank $2|1-k|$.  This tensor can be constructed from the $\mathbf{p}$ vectors by generalizing Eq.~(\ref{tfromp}).

The general relationship between the director field and the defect orientation tensor can be seen by expanding the tensor $\mathbf{nn}$.  From Eqs.~(\ref{thetaarounddefect}) and~(\ref{psidefinition}), an explicit calculation gives
\begin{align}
n_x n_x &= 1 - n_y n_y = \cos^2 \theta \\
&= \frac{1+\cos(2k\phi)\cos(2(1-k)\psi)-\sin(2k\phi)\sin(2(1-k)\psi)}{2},\nonumber\\
n_x n_y &= n_y n_x = \cos\theta\sin\theta \\
&= \frac{\sin(2k\phi)\cos(2(1-k)\psi)+\cos(2k\phi)\sin(2(1-k)\psi)}{2}.\nonumber
\end{align}
Here, each factor of $\cos(2k\phi)$ or $\sin(2k\phi)$ can be expressed in terms of $2|k|$ factors of $\mathbf{r}/|\mathbf{r}|$.  Likewise, each factor of $\cos(2{(1-k)}\psi)$ or $\sin(2{(1-k)}\psi)$ can be expressed in terms of $2|1-k|$ factors of $\mathbf{p}$, or equivalently in terms of a defect orientation tensor with rank ${2|1-k|}$.  A covariant, single-valued description of the defect orientation requires a tensor of that rank.

One important special case is a defect of topological charge $k=+1$, as shown in Fig.~\ref{nematicdefects}(e).  In that case, the tensor rank is $2|1-k|=0$, and hence the tensor is just a scalar.  This result is physically reasonable, because the defect is an isotropic object, which has no special directions going outward from the core.  Like other defect charges, the $+1$ defect has a parameter $\theta_0$ in the director field.  However, the significance of this parameter is different for a $+1$ defect than for other defect charges.  For a $+1$ defect, the parameter $\theta_0$ determines whether the director field points radially (splay deformation), tangentially (bend deformation), or somewhere in between.  This is a scalar property of the defect, which does not change as one moves around the defect core.  It is not related to an orientation of the defect.

\subsection{Generalization to $n$-atic phases}

Apart from the 2D nematic phase, researchers often consider phases with other types of orientational order.  If the orientational order parameter has $n$-fold symmetry, then the phase is called \hbox{$n$-atic}.  The case $n=1$ is a polar phase, such as a ferromagnet or ferroelectric, with a vector order parameter.  The case $n=2$ is a nematic phase, as discussed above.  The case $n=6$ is a hexatic phase, which commonly arises from 2D bond-orientational order.\cite{Nelson1979}

We can generalize the theory presented in this section to an $n$-atic phase with arbitrary $n$.  The elastic free energy still has the form of Eq.~(\ref{frankfreeenergy}), and the orientational order around a defect is still given by Eq.~(\ref{thetaarounddefect}).  For general $n$, the topological charge $k$ must be a positive or negative integer multiple of $1/n$, so that the director rotates through an angle of $2\pi/n$ around a loop about the defect.  The special directions where the orientational order matches the outward radial alignment are now given by
\begin{equation}
\theta=\phi\quad(\text{mod }2\pi/n).
\label{thetaequalsphinatic}
\end{equation}
Solving Eqs.~(\ref{thetaarounddefect}) and~(\ref{thetaequalsphinatic}) simultaneously, we find that these special radial directions occur at $\theta=\phi=\psi$, where
\begin{equation}
\psi=\frac{\theta_0}{1-k} \quad\left( \text{mod } \frac{2\pi}{n|1-k|} \right).
\label{psidefinitionnatic}
\end{equation}
Once again, we can construct the vector $\mathbf{p}=(\cos\psi,\sin\psi)$, which is now defined up to rotations through $2\pi/(n|1-k|)$.  Hence, there are $n|1-k|$ rotationally equivalent $\mathbf{p}$ vectors; i.e. the defect is an object with $n|1-k|$-fold rotational symmetry.  It can therefore be represented by a completely symmetric tensor of rank $n|1-k|$.

\begin{figure}
\centering
  \begin{tabular*}{0.5\textwidth}{@{\extracolsep{\fill}}cc}
    \includegraphics[width=4.2cm]{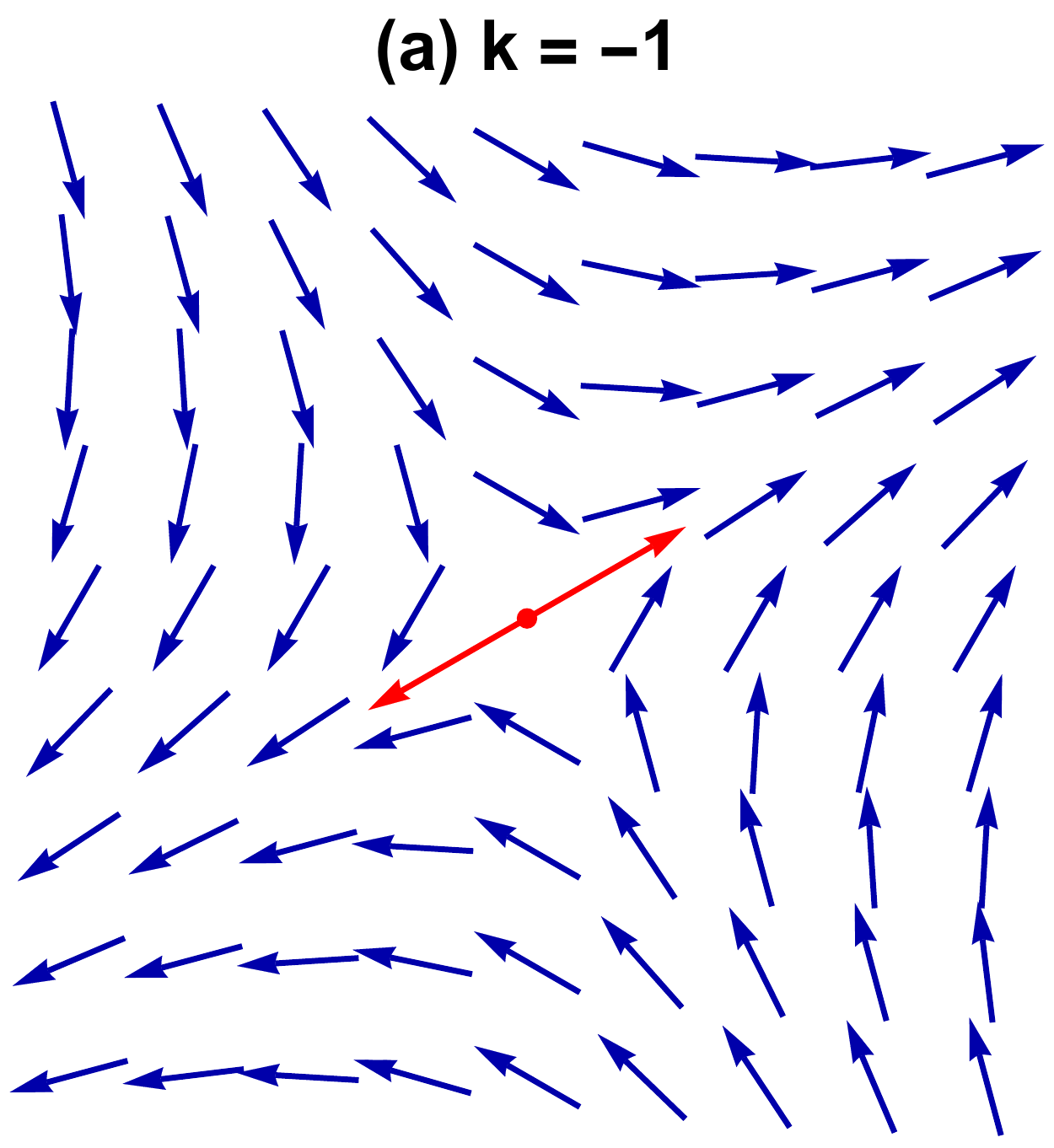} & \includegraphics[width=4.2cm]{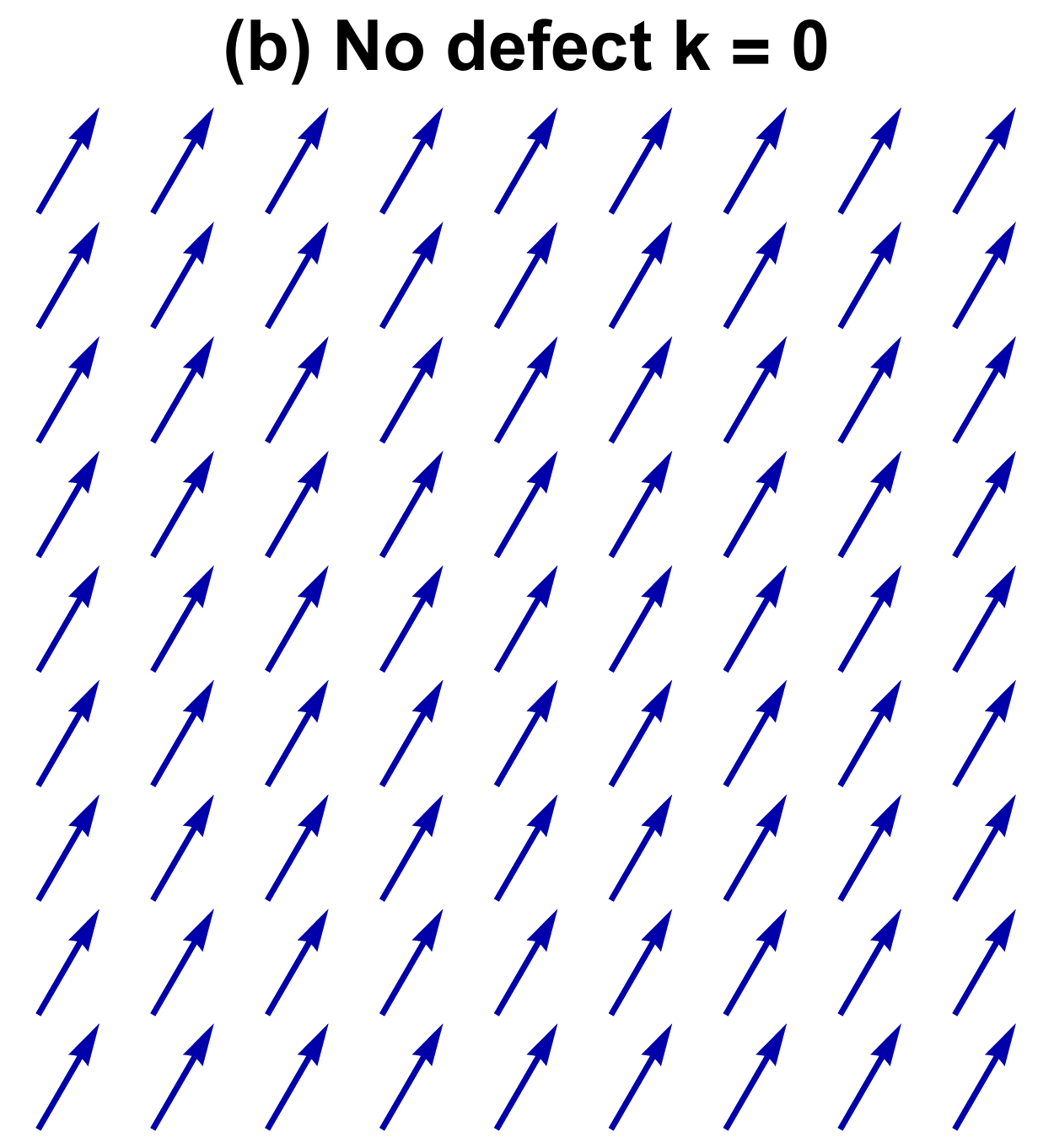} \\
    \includegraphics[width=4.2cm]{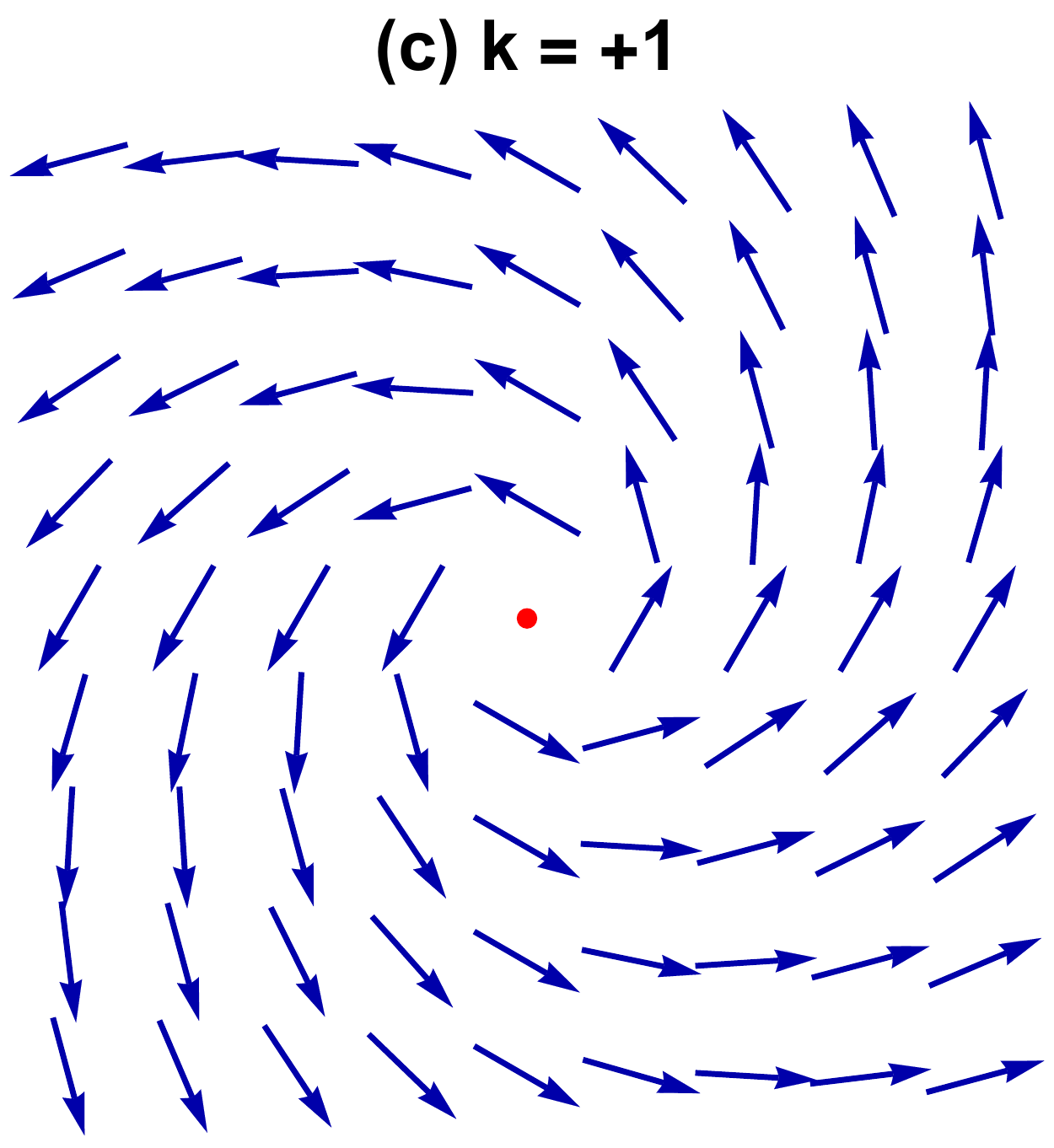} & \includegraphics[width=4.2cm]{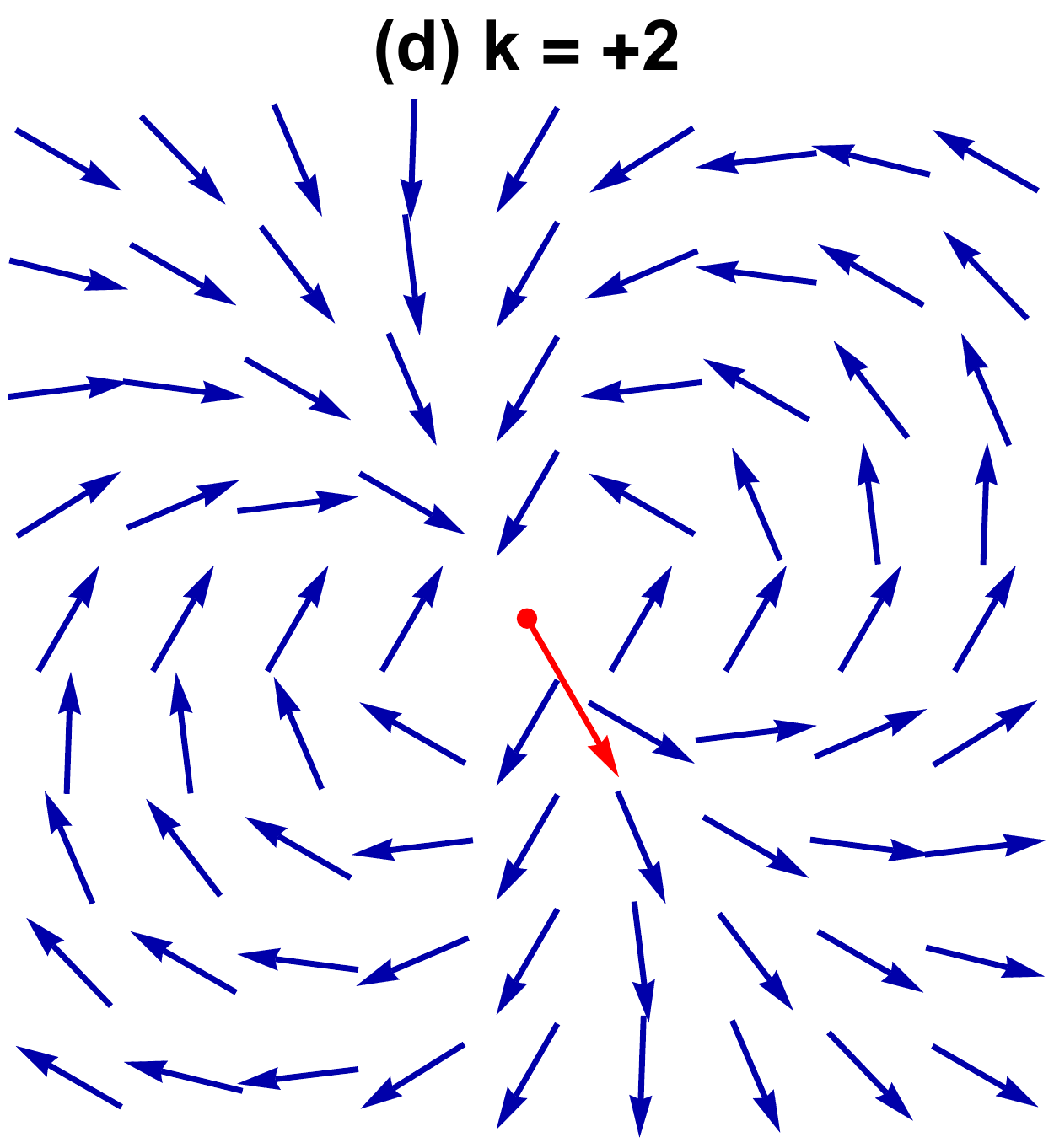} \\
  \end{tabular*}
  \caption{Examples of defects in a 2D polar phase ($n$-atic with $n=1$), with red arrows indicating the defect orientation.}
  \label{polardefects}
\end{figure}

Figure~\ref{polardefects} shows the example of a polar phase, which is $n$-atic with $n=1$.  The arrows can represent the magnetization in a ferromagnet, the electrostatic polarization in a ferroelectric, or the tilt in a smectic-C liquid crystal.  Here, the topological charge $k$ must be an integer.  For $k=+1$, the defect is characterized by a scalar (just as it is for the nematic case, or for any arbitrary $n$).  Indeed, the example of Fig.~\ref{polardefects}(c) shows that this defect appears the same in all directions.

For $k=-1$, the defect has 2-fold rotational symmetry, as indicated by the double-headed red arrow in Fig.~\ref{polardefects}(a).  The defect orientation is characterized by a tensor $\mathbf{T}$ of rank 2, with components
\begin{equation}
T_{ij}=\frac{1}{\sqrt{2}}\left(p^{(1)}_i p^{(1)}_j + p^{(2)}_i p^{(2)}_j - \delta_{ij} \right),
\end{equation}
with the $\delta_{ij}$ subtracted in order to make the tensor traceless.  This tensor is invariant under rotations of $\pi$, and it is normalized so that
$|\mathbf{T}|^2 \equiv T_{ij} T_{ij} = 1$.  To determine the vector field $\mathbf{n}$ around the defect, we construct the covariant expression
\begin{equation}
\mathbf{n} = \frac{\sqrt{2}\mathbf{T}\cdot\mathbf{r}}{|\mathbf{r}|},
\label{minusonecovariant}
\end{equation}
which is equivalent to Eq.~(\ref{thetaarounddefect}) with $k=-1$.
To determine $\mathbf{T}$ from $\mathbf{n}$, we take the gradient of Eq.~(\ref{minusonecovariant}),
\begin{equation}
\partial_j n_i = \frac{\sqrt{2}}{|\mathbf{r}|}\left[T_{ij}-\frac{T_{ik} r_k r_j}{|\mathbf{r}|^2}\right].
\end{equation}
By averaging this expression in a region around the defect, so that $\langle r_k r_j /|\mathbf{r}|^2\rangle = \delta_{kj}/2$, we obtain
\begin{equation}
\langle\partial_j n_i \rangle = \frac{1}{\sqrt{2}|\mathbf{r}|}T_{ij}.
\end{equation}
Hence, $\mathbf{T}$ is the normalized average
\begin{equation}
\mathbf{T}=\frac{\langle\nabla\mathbf{n}\rangle}{|\langle\nabla\mathbf{n}\rangle|},
\end{equation}
or more symmetrically,
\begin{equation}
T_{ij}=\frac{\langle\partial_i n_j +\partial_j n_i \rangle}{\sqrt{\langle\partial_a n_b +\partial_b n_a \rangle\langle\partial_a n_b +\partial_b n_a \rangle}}.
\end{equation}
The $\mathbf{p}$ vectors are $\pm1$ times the eigenvector of $\mathbf{T}$ with positive eigenvalue.

\section{Interaction of defects}

In this section, we determine the director field around two defects of arbitrary orientation in a 2D nematic phase.  We then use that director field to calculate the interaction energy as a function of the distance and relative orientation between the defects.

Suppose we have a defect of charge $-1/2$ at the origin, and a defect of charge $+1/2$ at the position $(R,0)$, assuming $R>0$.  We would like to specify the orientations of these two defects independently, so that
\begin{equation}
\theta(\mathbf{r})\approx
\begin{cases}
-\frac{1}{2}\tan^{-1}(y/x)     + \theta_1 & \text{for }\mathbf{r}\text{ near }(0,0), \\
+\frac{1}{2}\tan^{-1}(y/(x-R)) + \theta_2 & \text{for }\mathbf{r}\text{ near }(R,0).
\end{cases}
\end{equation}
Unfortunately, we \textit{cannot} just add these two solutions to obtain the solution for $\theta(\mathbf{r})$ everywhere.  This sum would have defects with the correct charges in the correct locations, but it would not have the correct defect orientations.  Rather, the constant terms $\theta_1$ and $\theta_2$ would just combine to give an overall constant, which would not allow us to fix the relative orientation of the two defects.

To find the director field around these defects, we must solve a differential equation with the appropriate boundary conditions.  The Euler-Lagrange equation associated with the Frank free energy~(\ref{frankfreeenergy}) is just Laplace's equation
\begin{equation}
\nabla^2 \theta = 0.
\end{equation}
Each defect has some core radius $r_\text{core}$, such that the nematic order is disrupted inside the core.  We apply boundary conditions at the core radius $r_\text{core}$ around each of the defects.  Hence, we must solve Laplace's equation on the full $(x,y)$ plane \textit{except} the two defect cores.

\begin{figure}
  \centering
  \includegraphics[width=8.8cm]{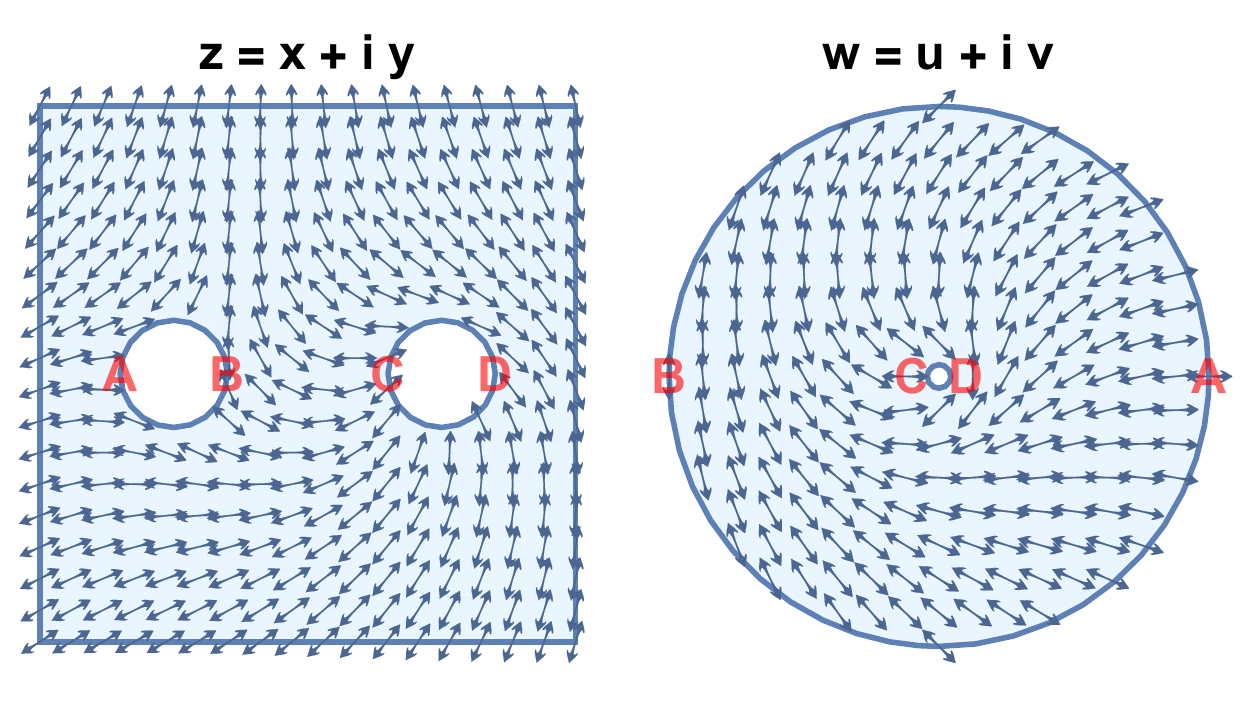}
  \caption{Solution of Laplace's equation for $\theta(\mathbf{r})$ by conformal mapping between the complex planes $z=x+iy$ and $w=u+iv$.  The $z$ domain goes out to infinity in all directions; it does not end at the square edge.  Red letters indicate corresponding points on the boundaries.  This example has $\theta_1 = \pi/2$ and $\theta_2 = 3\pi/4$.}
  \label{conformalmappingfigure}
\end{figure}

For this solution, we use the technique of conformal mapping, as illustrated in Fig.~\ref{conformalmappingfigure}.  We make a conformal transformation from the complex plane of $z=x+iy$ to the complex plane of $w=u+iv$, where
\begin{equation}
w = \frac{z-\gamma r_\text{core}}{\gamma z - r_\text{core}}, \qquad z = \frac{r_\text{core} (w-\gamma)}{\gamma w - 1},
\end{equation}
with
\begin{equation}
\gamma=\frac{R+\sqrt{R^2 - 4r_\text{core}^2}}{2 r_\text{core}}.
\end{equation}
With this transformation, the circular boundary of radius $r_\text{core}$ about $z=0$ maps onto a circular boundary of radius $w_\text{max}=1$ about $w=0$, and the circular boundary of radius $r_\text{core}$ about $z=R$ maps onto a circular boundary of radius
\begin{equation}
w_\text{min}=\frac{R^2 - 2 r_\text{core}^2 - R\sqrt{R^2 - 4 r_\text{core}^2}}{2r_\text{core}^2},
\end{equation}
also about $w=0$.  Hence, we must solve Laplace's equation between two concentric circles in the complex $w$ plane.  The solution is
\begin{equation}
\theta(w)=\frac{\operatorname{Im}(\log w)}{2}
+\frac{\delta\theta \operatorname{Re}(\log w)}{\log w_\text{min}}
+\Theta,
\end{equation}
where
\begin{equation}
\delta\theta = \theta_2 - \theta_1 + \frac{\pi}{2}, \qquad \Theta=\theta_1 - \frac{\pi}{2}.
\end{equation}
Transforming back into the $(x,y)$ plane, this solution becomes
\begin{align}
\theta(\mathbf{r})=&-\frac{1}{2}\tan^{-1}\left(\frac{y}{x+r_\text{core}\gamma-R}\right)+\frac{1}{2}\tan^{-1}\left(\frac{y}{x-r_\text{core}\gamma}\right)\\
&+\frac{\delta\theta}{2\log w_\text{min}}\log\left(\frac{y^2+(x-r_\text{core}\gamma)^2}{y^2 \gamma^2 + (r_\text{core}-x\gamma)^2}\right)
+\Theta.\nonumber
\end{align}
In the limit of small core radius $r_\text{core}$, it can be approximated by
\begin{align}
\label{minushalfplushalf}
\theta(\mathbf{r})=&-\frac{1}{2}\tan^{-1}\left(\frac{y}{x}\right)+\frac{1}{2}\tan^{-1}\left(\frac{y}{x-R}\right)\\
&+\frac{\delta\theta}{2}
\left[1+\frac{\log(x^2+y^2)-\log((x-R)^2+y^2)}{\log(R^2)-\log(r_\text{core}^2)}\right]
+\Theta.\nonumber
\end{align}
Here, the inverse tangents and the additive constant are the usual expression for the director field around two defects.  The term proportional to $\delta\theta$ is a new term, which is required to specify the relative orientation of the two defects.

\begin{figure}
\centering
\includegraphics[width=8.8cm]{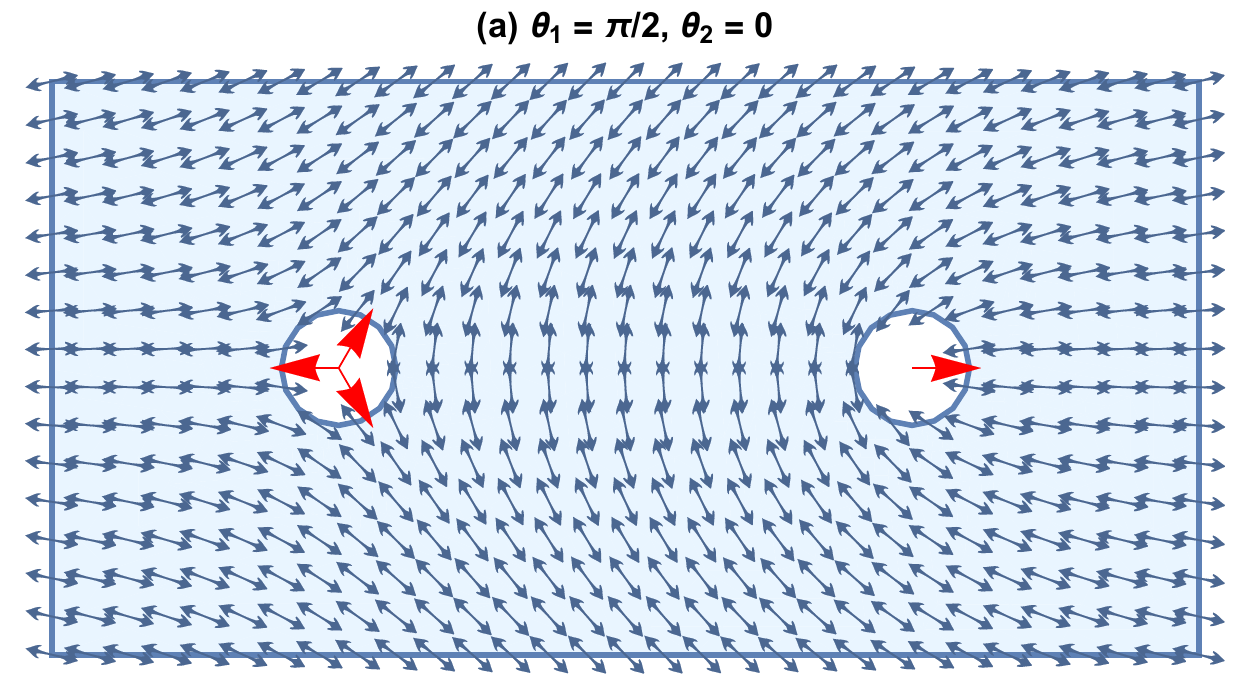}\\
\includegraphics[width=8.8cm]{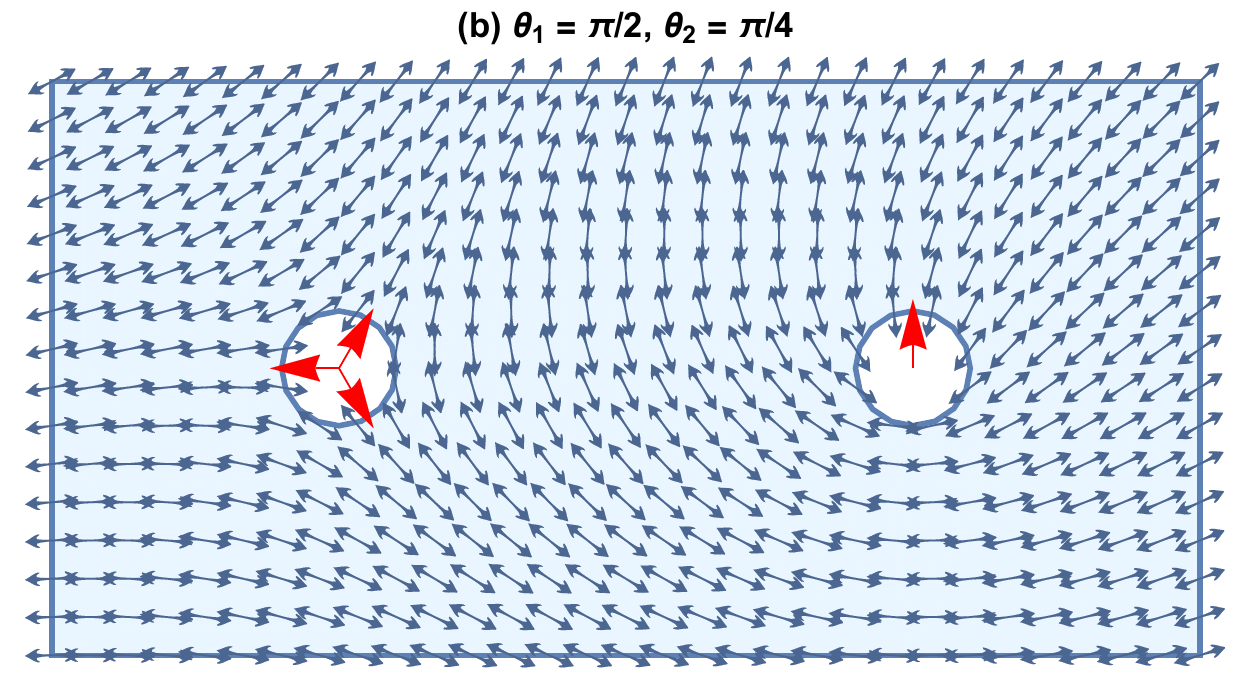}\\
\includegraphics[width=8.8cm]{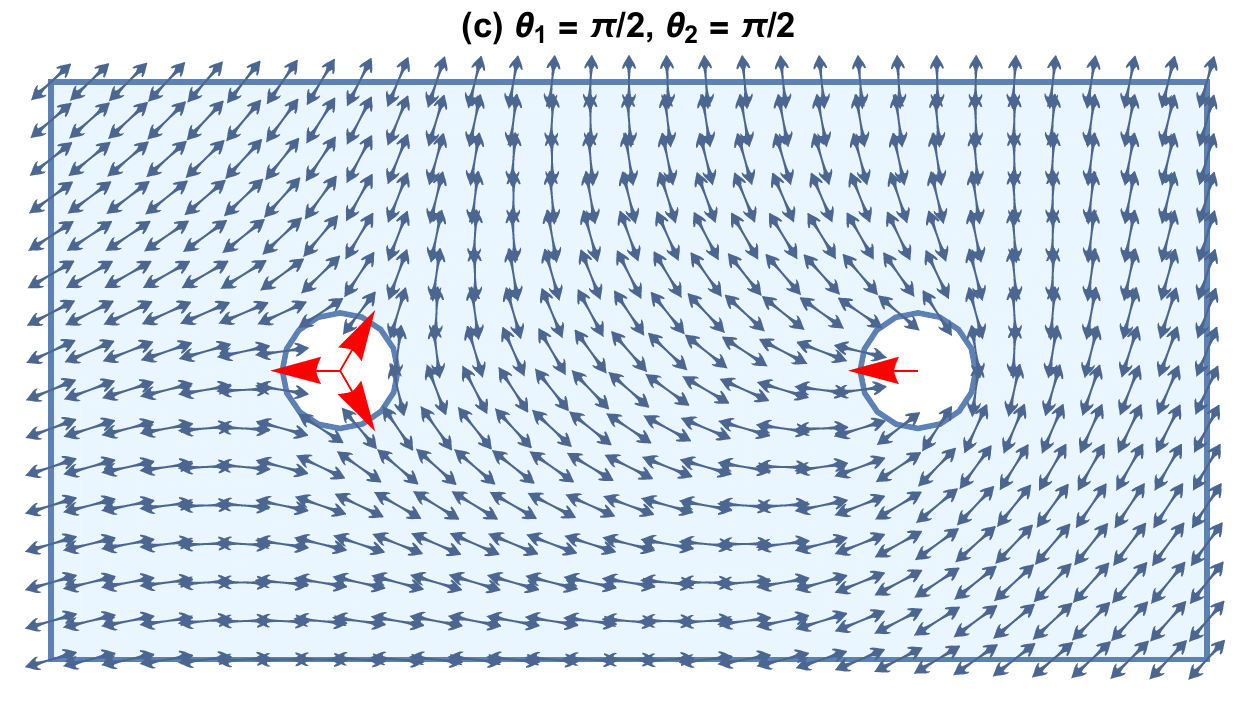}\\
\includegraphics[width=8.8cm]{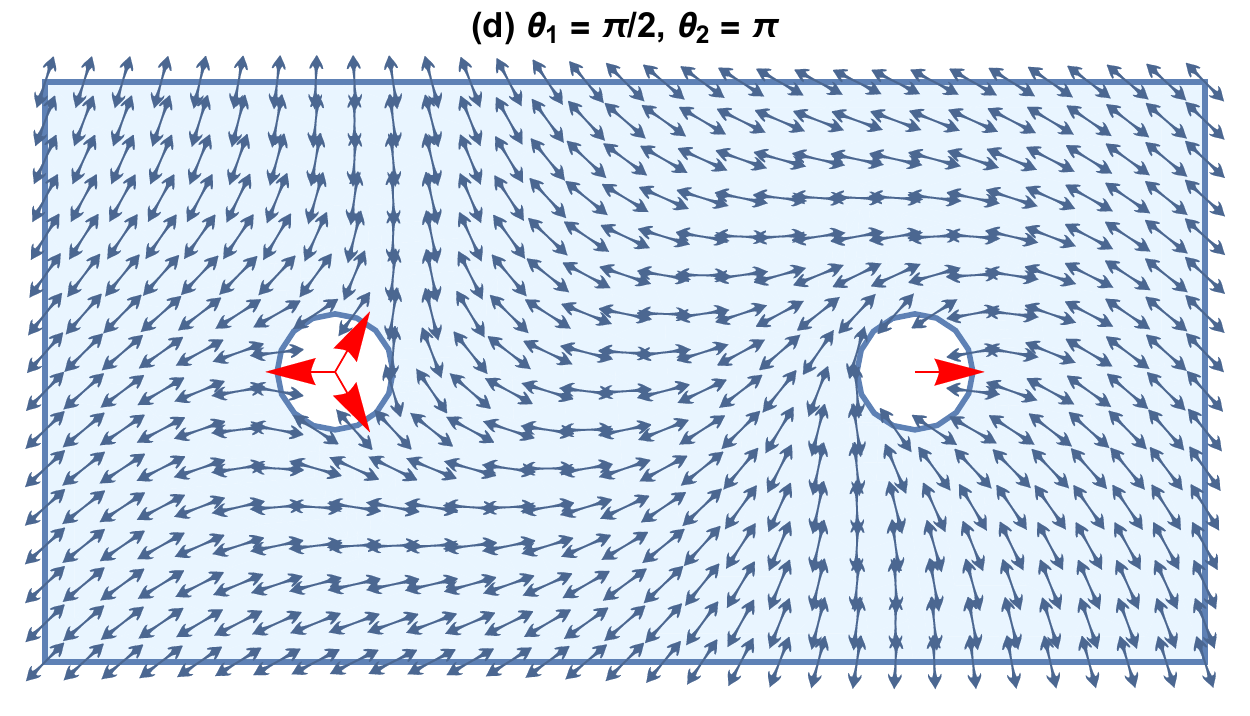}
\caption{Director field around two interacting defects of topological charges $k_1 = -1/2$ and $k_2 = +1/2$.  The first defect has a fixed orientation, while the second defect rotates through a full circle.  The white circles indicate the defect cores, and the red arrows indicate the defect orientations.}
\label{interactingdefects}
\end{figure}

Figure~\ref{interactingdefects} shows the director field of Eq.~(\ref{minushalfplushalf}), in the case where $r_\text{core}/R=0.1$.  The $-1/2$ defect on the left has the fixed orientation $\theta_1 = \pi/2$, which implies $\psi_1 = \pi/3$ by the argument in the previous section.  The $+1/2$ defect on the right has an orientation that rotates from $\theta_2 = 0$ to $\pi$, which implies $\psi_2=0$ to $2\pi$.  In Fig.~\ref{interactingdefects}(a), the defects clearly have the optimal relative orientation, and the director field has the usual form as the sum of inverse tangents.  As the $+1/2$ defect rotates, the director field becomes more distorted.  When the defect rotates through a full circle, the texture does not return to its original form, because extra distortion has wound up throughout the director field.

Note that we can specify the defect orientations at the specific core radius $r_\text{core}$.  We cannot specify the orientations at the centers of the defects, because the problem becomes mathematically undefined in the limit of $r_\text{core}\to0$.

We can generalize the form of Eq.~(\ref{minushalfplushalf}) to describe arbitrary defect charges $k_1$ and $k_2$ at arbitrary positions $\mathbf{R}_1=(x_1,y_1)$ and $\mathbf{R}_1=(x_2,y_2)$.  This generalization gives
\begin{align}
\label{k1k2}
\theta(\mathbf{r})=&k_1\tan^{-1}\left(\frac{y-y_1}{x-x_1}\right)+k_2\tan^{-1}\left(\frac{y-y_2}{x-x_2}\right)\\
&+\frac{\delta\theta}{2}
\left[1+\frac{\log(|\mathbf{r}-\mathbf{R}_1 |^2)-\log(|\mathbf{r}-\mathbf{R}_2 |^2)}{\log(|\mathbf{R}_1 -\mathbf{R}_2 |^2)-\log(r_\text{core}^2)}\right]
+\Theta,\nonumber
\end{align}
where
\begin{align}
\delta\theta =& \theta_2 - \theta_1 + k_2 \tan^{-1}\left(\frac{y_1-y_2}{x_1-x_2}\right) - k_1 \tan^{-1}\left(\frac{y_2-y_1}{x_2-x_1}\right),\nonumber\\
\Theta =&  \theta_1 - k_2 \tan^{-1}\left(\frac{y_1-y_2}{x_1-x_2}\right).
\end{align}
The visualization of the director field shows the same type of behavior as in Fig.~\ref{interactingdefects}.  There is an optimal director field when the relative orientation has $\delta\theta=0$, and the configuration becomes more distorted as $\delta\theta$ increases.

To calculate the elastic free energy associated with the distorted director field, we put the expression for $\theta(\mathbf{r})$ from Eq.~(\ref{k1k2}) into the Frank free energy of Eq.~(\ref{frankfreeenergy}), and integrate over the plane out to the system size of $R_\text{max}$.  The integral is done in Mathematica, using Cartesian coordinates such that $x_1 = -\frac{1}{2}R$, $x_2 = \frac{1}{2}R$, and $y_1 = y_2 = 0$, over the domain $-R_\text{max} < x < -\frac{1}{2}R - a$, $-\frac{1}{2}R + a < x < \frac{1}{2}R -a$, and $\frac{1}{2}R + a < x < R_\text{max}$, with $-\infty < y < \infty$.  The result is
\begin{align}
F=&\pi K (k_1 + k_2)^2 \log\left(\frac{R_\text{max}}{r_\text{core}}\right) 
- 2\pi K k_1 k_2 \log\left(\frac{|\mathbf{R}_1 -\mathbf{R}_2 |}{2r_\text{core}}\right)\nonumber\\
&+ \frac{\pi K \delta\theta^2}{2} \frac{\log(|\mathbf{R}_1 -\mathbf{R}_2 |/(2r_\text{core}))}{[\log(|\mathbf{R}_1 -\mathbf{R}_2 |/r_\text{core})]^2} .
\label{defectinteraction}
\end{align}
Here, the first term is the usual energy cost of a defect pair with net topological charge $(k_1 + k_2 )$, which diverges logarithmically with system size $R_\text{max}$ unless the net charge is zero.  The second term is the usual Coulomb-like logarithmic interaction between two defects, which is repulsive for like charges and attractive for opposite charges.  The third term is a new contribution, which favors orientational alignment between the defects toward the optimal orientation of $\delta\theta=0$.  It creates an aligning torque
\begin{equation}
-\frac{\partial F}{\partial(\delta\theta)}=
-\pi K \frac{\log(|\mathbf{R}_1 -\mathbf{R}_2 |/(2r_\text{core}))}{[\log(|\mathbf{R}_1 -\mathbf{R}_2 |/r_\text{core})]^2} \delta\theta,
\end{equation}
and this torque decreases as $\pi K/\log(|\mathbf{R}_1 -\mathbf{R}_2 |/r_\text{core})$ when the defect separation is much greater than the core radius.

In Eq.~(\ref{defectinteraction}), the orientational interaction is expressed in terms of $\delta\theta$ rather than in terms of the defect orientation vectors $\mathbf{p}$ or tensors $\mathbf{T}$ defined in the previous section.  This is necessary because the interaction is not a single-valued function of $\mathbf{p}$ or $\mathbf{T}$.  As an example, the textures in Figs.~\ref{interactingdefects}(a) and~\ref{interactingdefects}(d) have the same $\mathbf{p}$ and $\mathbf{T}$, but clearly the texture in Fig.~\ref{interactingdefects}(d) is more distorted and has a higher elastic free energy.  When $\delta\theta$ is \textit{small}, it may be possible to express the orientational interaction in terms of $\mathbf{p}$ or $\mathbf{T}$.  However, that cannot work when $\delta\theta$ is large and the texture is wound up, as in Fig.~\ref{interactingdefects}(d).

Our defect interaction of Eq.~(\ref{defectinteraction}) can be compared with the work of Vromans and Giomi.~\cite{Vromans2016}  They calculate the defect interaction in two ways.  In their first method, they construct a field $\theta(\mathbf{r})$ that linearly interpolates between the arctangents around the two defects, and then calculate the free energy associated with this field.  For two $+1/2$ disclinations separated by a distance $d$ in the $x$ direction, in a square $L\times L$ domain, they find an orientation-dependent part of the free energy that scales as $(K L/d)(\delta\psi-\pi)^2$ (where $\delta\psi-\pi$ is $2\delta\theta$ in our notation).  One should note that their linear interpolation is \emph{not} a minimizer of the free energy, and hence the free energy that they calculate is higher than our free energy, with a different dependence on system size and defect separation.  We would argue that the interaction between defects is only defined when the director field between the defects is a minimizer of the free energy (or perhaps has Casimir-like fluctuations about the minimizer).  To our understanding, the free energy that they calculate is the free energy of a particular choice of director field, but it cannot be considered as a property of the defects.

In their second method, Vromans and Giomi use an image construction to model like-sign defects with charge $k_1 = k_2 \equiv k$, in the limit of large system size $L$ and large defect separation $d$, and find the orientation-dependent part of the interaction as $F=-\pi K k^2 \log(1-\mathbf{p}_1 \cdot \mathbf{p}_2 )$.  Although this result is expressed in terms of the vectors $\mathbf{p}_1$ and $\mathbf{p}_2$, it really only applies over a limited domain of $0<\delta\psi<2\pi$ around the minimum; it does not describe arbitrary windings of $\mathbf{p}_1$ or $\mathbf{p}_2$ through a full circle.  The quadratic minimum of this function is similar to our defect interaction, but it shows deviations from quadratic behavior that we do not find, and it does not show the logarithmic decay of our interaction.  

We have done numerical simulations to check the result of Eq.~(\ref{defectinteraction}).  In these simulations, we construct a 2D hexagonal lattice of spins interacting through the energy $E=J\sum_{\langle i,j\rangle} {[1-\cos(2(\theta_i - \theta_j))]}$, which is a discretized approximation to the Frank free energy of Eq.~(\ref{frankfreeenergy}).  We fix the position and orientations of two defects by placing effective ``particles'' on certain plaquettes between lattice sites.  Each particle has strong anchoring to fix the spins on its boundary in a defect configuration, with topological charge of $\pm1/2$, and with specified orientation.  We then use a relaxation method to minimize the total energy over the spins on all of the non-anchored sites.  Through this method, we find the minimum total energy as a function of relative orientation.  The results are consistent with the predicted quadratic dependence on $\delta\theta$.  We were not able to check the distance dependence of the orientational interaction, because the logarithmic decay is very slow in comparison with accessible length scales.

\section{Motion of defects}

In this section, we investigate the motion of two opposite-charged defects as they annihilate each other, to determine how the motion depends on defect orientation.

In order to describe a system in which the defects are free to move, we generalize the theory to represent the magnitude and direction of nematic order by a tensor
\begin{align}
\mathbf{Q}(\mathbf{r},t) =& S(\mathbf{r},t)\left[2\mathbf{n}(\mathbf{r},t)\mathbf{n}(\mathbf{r},t)-\mathbf{I}\right]\\
=& S(\mathbf{r},t)
\begin{pmatrix}
\cos2\theta(\mathbf{r},t) &  \sin2\theta(\mathbf{r},t) \\
\sin2\theta(\mathbf{r},t) & -\cos2\theta(\mathbf{r},t)
\end{pmatrix}
.\nonumber
\end{align}
In that formalism, the free energy can be expressed as
\begin{align}
F=& \int d^2 r \left[-\frac{a}{4} Tr(\mathbf{Q}^2) + \frac{b}{16} (Tr(\mathbf{Q}^2))^2 
+ \frac{L}{4} (\partial_i Q_{jk})(\partial_i Q_{jk})\right]\nonumber\\
=& \int d^2 r \left[-\frac{a}{2} S^2 + \frac{b}{4} S^4 + \frac{L}{2}|\nabla S|^2 + 2 L S^2 |\nabla\theta|^2 \right].
\end{align}
Away from defects, where gradients of $\theta$ are small, the bulk value of the scalar order parameter is $S=\sqrt{a/b}$.  At the defect points, where $\theta$ is singular, the scalar order parameter goes to $S=0$.  In each core around a defect point, the scalar order parameter varies over a length scale $r_\text{core}=\sqrt{L/a}$.

To model the time evolution of nematic order, we use the equations for pure relaxational dynamics
\begin{equation}
\frac{\partial Q_{xx}(\mathbf{r},t)}{\partial t} = -\frac{1}{\gamma_1} \frac{\delta F}{\delta Q_{xx}(\mathbf{r},t)}, \quad
\frac{\partial Q_{xy}(\mathbf{r},t)}{\partial t} = -\frac{1}{\gamma_1} \frac{\delta F}{\delta Q_{xy}(\mathbf{r},t)},
\end{equation}
where $\gamma_1$ is the rotational viscosity, and $\mathbf{Q}$ is constrained to be symmetric and traceless.  For the initial condition, we use a director field containing two defects, with topological charges $k_1 = -1/2$ and $k_2 = +1/2$ and arbitrary initial orientations $\theta_1$ and $\theta_2$, as found in Eq.~(\ref{k1k2}).  We assume that the initial $S(\mathbf{r},t)$ has the bulk value everywhere except in the defect cores, with the functional form
\begin{equation}
S_\text{initial}(\mathbf{r},t)=\frac{2\sqrt{a/b}}{\sqrt{1+r_\text{core}^2/|\mathbf{r}-\mathbf{R}_1 |^2}+
\sqrt{1+r_\text{core}^2/|\mathbf{r}-\mathbf{R}_2 |^2}}.
\end{equation}
This expression is physically motivated, in that it goes to the bulk value of $S=\sqrt{a/b}$ away from the defects, and it goes to zero linearly at the defects, but the exact form is arbitrary.  We use open boundary conditions, at which $\mathbf{Q}$ is free and the normal derivatives vanish.  We solve the differential equations in Mathematica, iterating forward in time until the defects annihilate each other.  At each time, we find the defect positions by searching for points where $S(\mathbf{r},t)$ vanishes, and then find the defect orientations by the procedure described in Sec.~2.

\begin{figure}
\centering
\includegraphics[width=8.8cm]{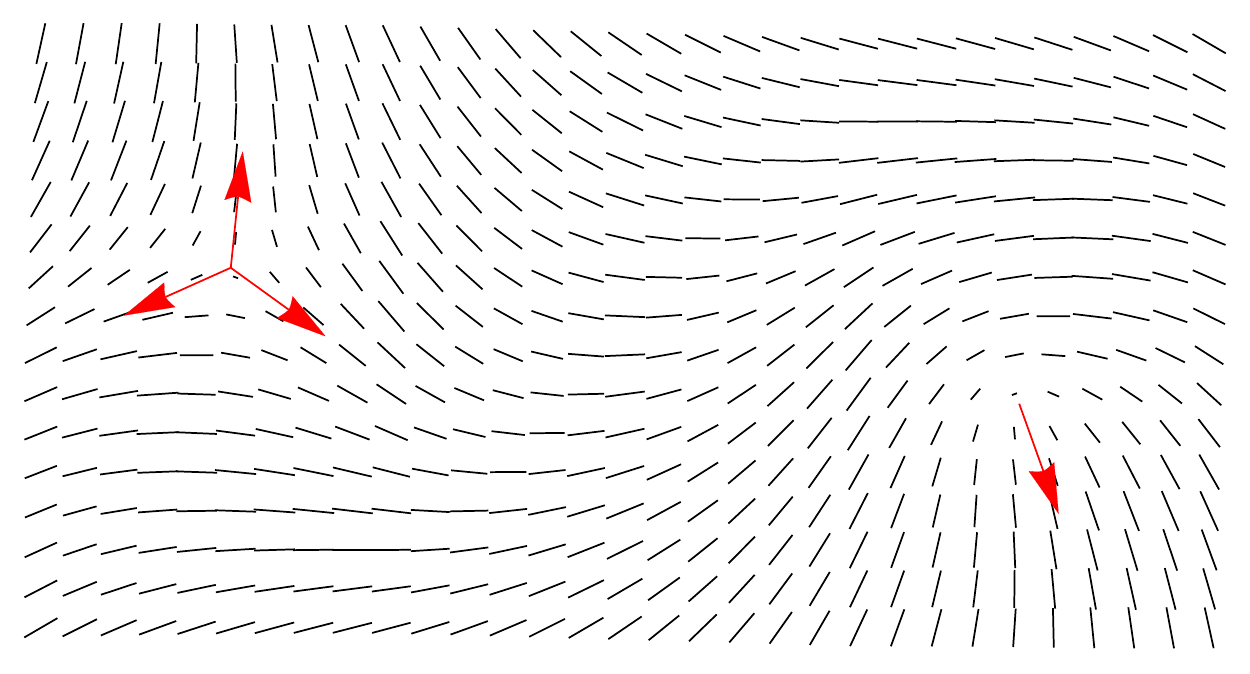}\\
\includegraphics[width=8.8cm]{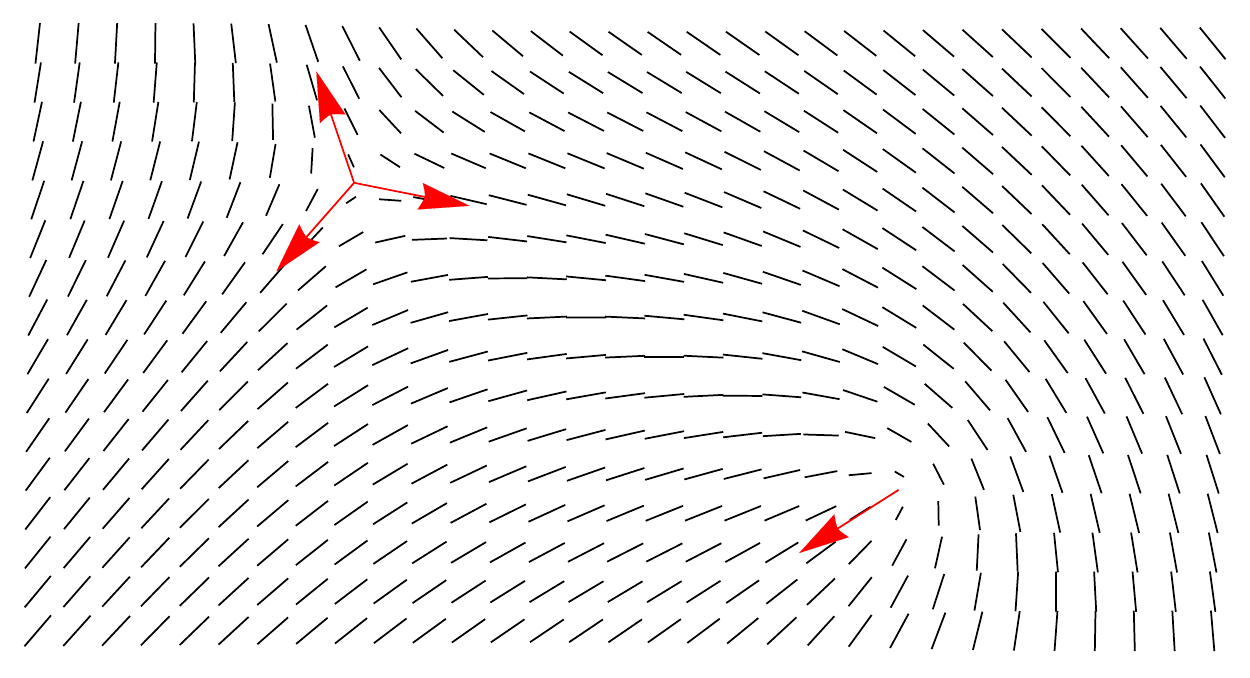}\\
\includegraphics[width=8.8cm]{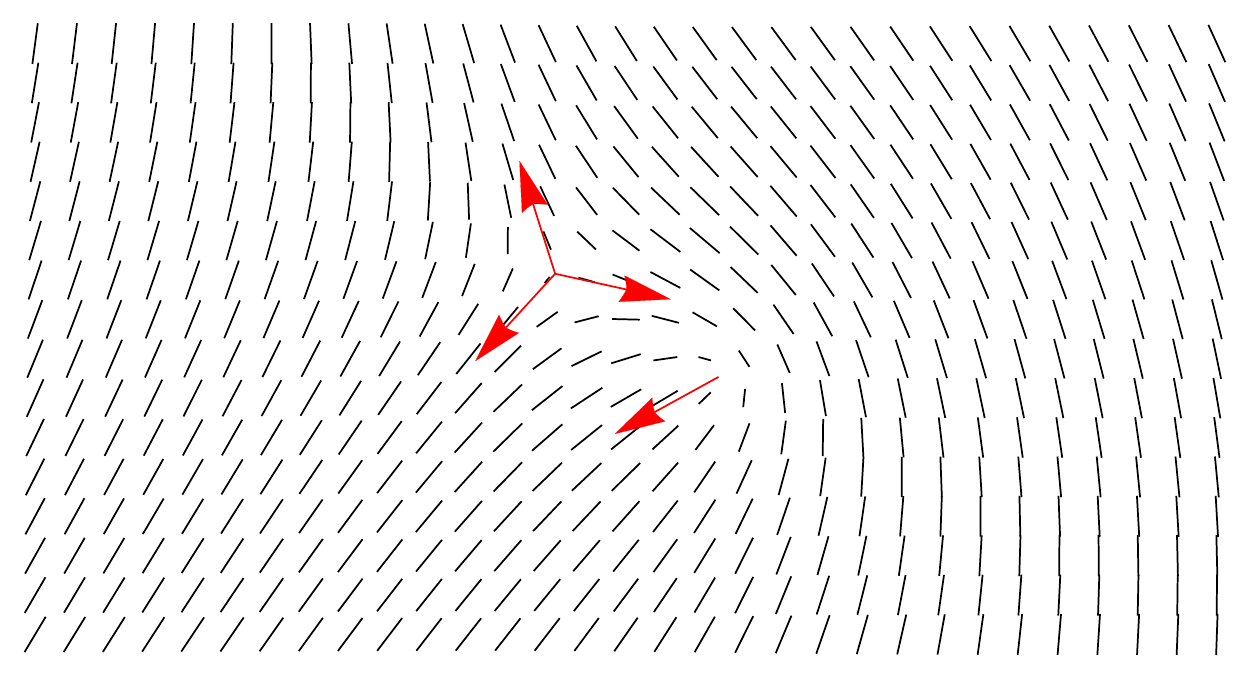}\\
\includegraphics[width=8.8cm]{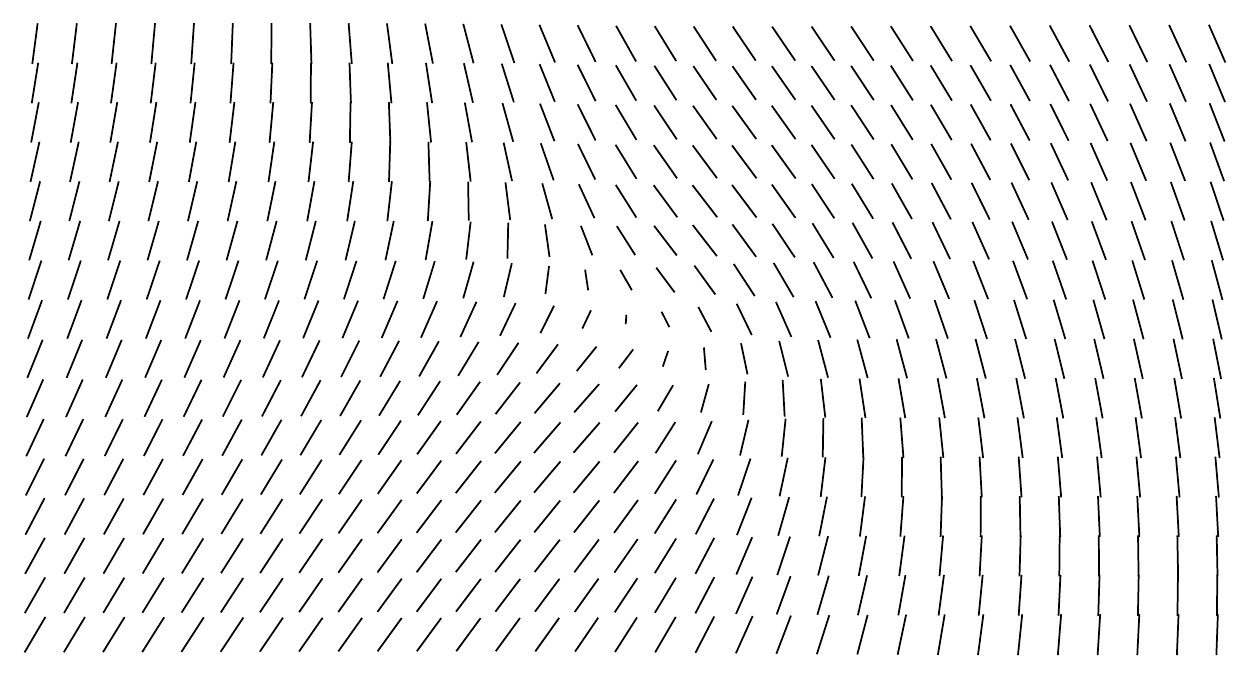}
\caption{Snapshots of the dynamic evolution of the liquid crystal order, beginning with two defects with topological charges of $-1/2$ and $+1/2$ at an unfavorable relative orientation, $\delta\theta=\pi$.}
\label{annihilatingdefects}
\end{figure}

Figure~\ref{annihilatingdefects} shows a series of snapshots of the time evolution of the $\mathbf{Q}$ tensor field.  The system begins with the two defects at an unfavorable relative orientation, with $\delta\theta=\pi$.  In the early stage of the dynamic process, the defects rapidly rotate into the optimal relative orientation.  While they rotate, they also move in the $y$ direction, which is transverse to their separation in the $x$ direction.  Once they reach the optimal relative orientation, the dynamics becomes much slower.  In this stage of the process, the defects move straight toward each other.  After they annihilate each other, they leave a defect-free configuration, which eventually becomes uniform.

\begin{figure}
\centering
\includegraphics[width=8.8cm]{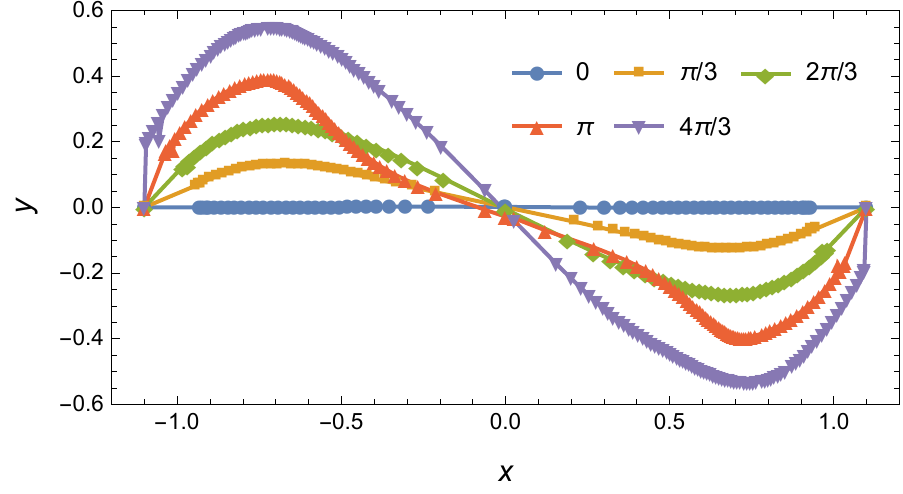}
\caption{Trajectories of the annihilating $-1/2$ and $+1/2$ defects, for several values of the initial relative orientation $\delta\theta$.}
\label{defecttrajectories}
\end{figure}

To demonstrate the influence of defect orientation, Fig.~\ref{defecttrajectories} shows the defect trajectories for several values of the initial $\delta\theta$.  In this figure, the symbols represent the positions at equally spaced times, and hence the spacing between the symbols indicates the defect velocity.  For initial $\delta\theta=0$, the defects are already at the optimal relative orientation at the beginning of the calculation.  In that case, they move straight toward each other.  The motion is initially rapid as the $\mathbf{Q}$ tensor field relaxes from a somewhat arbitrary initial configuration, then it slows down once the system reaches an almost-stable configuration with two defects, then it accelerates as the defects grow closer and the attractive force between them increases.  By contrast, for initial $\delta\theta\not=0$, the initial stage of motion involves both rotation and translation in the $y$ direction, transverse to the inter-defect separation, until the optimal relative orientation is reached.  For larger initial $\delta\theta$, the amount of translation is greater.  Once the defects have the optimal relative orientation, their motion becomes much slower, and then later accelerates after the defects grow closer.

We emphasize that the initial motion is in the $y$ direction, transverse to the inter-defect separation, in spite of the fact that the defect interaction of Eq.~(\ref{defectinteraction}) depends only on the magnitude of the separation.  In other words, the defects move in the transverse direction although the interaction provides no force in the transverse direction.  This behavior is an important feature of the motion of objects with internal orientation, which have an anisotropic drag as they move through a medium.  In this sense, the motion of a defect is analogous to the motion of a sailboat, which can move transverse to the wind because of the orientation of the boat and the sail.

Our results for defect motion are actually quite similar to the results of Vromans and Giomi.~\cite{Vromans2016}  They also find curved trajectories, which are induced by the initial relative orientation of the defects.  Their calculations of the dynamic evolution of the $\mathbf{Q}$ tensor are not affected by the issues involving the defect interaction discussed in the previous section.

\section{Discussion}

In this paper, we have examined the concept of defect orientation, which was initially developed by Vromans and Giomi.~\cite{Vromans2016}  Through this study, we partially agree and partially disagree with their work.  We agree with them about the vector description of defects with topological charge $+1/2$, and about the motion of defects.  We suggest that a tensor formalism provides a clearer way to describe defects with other topological charges, although it is consistent with their vector formalism.  We disagree with them about the interaction between defects.  Of course, despite these specific differences, we recognize their contribution of introducing this concept into the theoretical physics literature.

We must emphasize that defect orientation is not a topological invariant like defect charge.  Indeed, it is not a topological concept at all.  Rather, it is a geometric feature of defects, which will certainly change as a function of time.  In that respect, it is analogous to defect position, which also changes as a function of time.  Physicists often speak of defects as if they were effective ``particles,'' which can move around inside a liquid crystal.  We argue that they should be considered as particles with orientation as well as position.  They can exhibit both rotational and translational motion, and these two types of motion are coupled together.

We expect that the concept of defect orientation can be generalized in several ways.  One important generalization will be to connect it back to \textit{active} nematic liquid crystals.  While this concept was inspired by experiments and simulations on active materials, our calculations have so far only considered the case of equilibrium liquid crystals.  Once it is combined with theories of active nematics, there will certainly be a coupling between vector orientation of $+1/2$ defects and active motion, and there may also be new ways to understand the long-range ordering of defect orientation.  A further generalization will be to 3D nematic liquid crystals.  In general, 3D nematics have a more complex set of defects than 2D nematics, with both hedgehog points and disclination lines.  These defects have their own types of orientation, as investigated by the graphical visualizations and topological arguments of \v{C}opar et al.,\cite{Copar2011,Copar2014} and we expect that these orientations can be understood through new tensor constructions.  Finally, the concept of defect orientation can be connected with other aspects of liquid crystal theory, including backflow, interaction with colloidal particles, and background alignment of the director field, leading to new insights into defect behavior.

We would like to thank A.~Baskaran and L.~Giomi for helpful discussions.  This work was supported by National Science Foundation Grant No.~DMR-1409658. 





\bibliography{defectorientation} 
\bibliographystyle{rsc} 

\end{document}